\documentclass[onecolumn,showpacs,amsmath,amssymb,noshowpacs]
{revtex4-2}

\usepackage{array} 
\usepackage{braket}

\usepackage{tikz}
\usetikzlibrary{fadings}
\usetikzlibrary{patterns}
\usetikzlibrary{shadows.blur}
\usetikzlibrary{shapes}

\usepackage{definitions}%

\begin{document}

\title{Hall conductivity as the topological invariant in magnetic Brillouin zone in the presence of interactions}

\author{M. Selch}
\affiliation{Physics Department, Ariel University, Ariel 40700, Israel}

\author{M. Suleymanov}
\affiliation{Physics Department, Ariel University, Ariel 40700, Israel}

\author{C. X. Zhang}
\affiliation{Wuerzburg University, Am Hubland
	97074 Wuerzburg, Germany }

\author{M. A. Zubkov}
\affiliation{Physics Department, Ariel University, Ariel 40700, Israel}

\begin{abstract}
Hall conductivity for the intrinsic anomalous quantum Hall effect in homogeneous systems is given by the topological invariant composed of the Green function depending on momentum of quasiparticle. This expression reveals correspondence with the mathematical notion of the degree of mapping. A more involved situation takes place for the quantum Hall effect in the presence of external magnetic field. In this case the mentioned expression remains valid if the Green function is taken in a specific representation, where it becomes the infinite - dimensional matrix \cite{Imai:1990zz} or if it is replaced by its Wigner transformation while ordinary products are replaced by the Moyal products \cite{ZW2019}. Both these  expressions, unfortunately, are much more complicated and might be useless for the practical calculations. Here we represent the alternative representation for the Hall conductivity of a uniform system in the presence of constant magnetic field. The Hall conductivity is expressed through the Green function taken in Harper representation, when its nonhomogeneity is attributed to the matrix structure while functional dependence is on one momentum that belongs to magnetic Brillouin zone.  Our consideration for the interacting systems is non - perturbative and is  based on the Schwinger - Dyson equations truncated in a reasonable way. We demonstrate that in this approximation the expression for the Hall conductivity in Harper representation remains valid, where the interacting Green function is to be used instead of the non - interacting one. We, therefore, propose that the obtained expression may be used for the topological description of fractional quantum Hall effect.
\end{abstract}



\maketitle

\section{Introduction}


The first topological expression for the QHE conductivity has been proposed in \cite{TKNN} for the ideal two dimensional non-interacting condensed matter systems in the presence of constant external magnetic field. This expression is proportional to the TKNN invariant, which is the integral of Berry curvature in magnetic Brillouin zone
over the occupied electronic states \cite{Fradkin,Tong:2016kpv,Hatsugai}. This expression also remains valid for the description of integer intrinsic QHE in 2d topological insulators. Besides, this approach has been extended to the three space-dimensional (3d) topological insulators
(see, for example, \cite{Hall3DTI}). The introduction
of weak interactions as well as disorder (to the ideal homogeneous system of non - interacting particles) does not affect total conductivity of integer QHE, considered as a function of chemical potential at fixed value of magnetic field (for those values of the chemical potential that belong to the Hall plateaus). It is important, therefore, to express Hall conductivity
through the Green functions that are well - defined within interacting theory.
Such an expression for the conductivity of intrinsic anomalous QHE (AQHE) in $2+1$ D systems
has been given through the Green functions in \cite{Matsuyama:1986us,Imai:1990zz,Volovik0} (see also Chapter 21.2.1 in \cite{Volovik2003}).
The extension of such a construction to various 3d systems has also been proposed \cite{Z2016_1}. The resulting expression allows to describe the AQHE in Weyl semimetals \cite{semimetal_effects10,semimetal_effects11,semimetal_effects12,semimetal_effects13,Zyuzin:2012tv,tewary}. The similar topological invariants have also been discussed
in \cite{Gurarie2011,EssinGurarie2011}. The AQHE conductivity is given by the expressions of  \cite{Matsuyama:1986us,Volovik0,Volovik2003,Z2016_1} through the two - point Green functions also for the interacting systems. The general proof of this statement has been given in \cite{ZZ2019_0}.

In \cite{ZW2019} the construction of \cite{Matsuyama:1986us,Imai:1990zz,Volovik0} was extended to the essentially non - homogeneous systems. It appears that the Hall conductivity is expressed through the Wigner transformed two - point Green function.
This gives an alternative proof that disorder does not affect the total QHE conductivity (although the local Hall current is pushed by disorder towards the boundary of the sample). It is worth mentioning that the role of disorder
in QHE has been widely discussed in the past \cite{Fradkin,TKNN2,QHEr,Tong:2016kpv,Zheng+Ando2002}. The absence of corrections due to weak Coulomb interactions to the QHE ferromagnetic metal was discussed in \cite{nocorrectionstoQHE}. Inter - electron interactions and their relation to QHE have also been discussed long time ago (see, for example,  \cite{Altshuler0,Altshuler,Ishikawa:2003tu,Imai:1990zz}). The proof of the absence of radiative corrections to Hall conductivity (to all orders in perturbation theory) in the presence of external magnetic field has been given in \cite{ZZ2019_1,ZZ2021} (see also \cite{parity_anomaly}).

Unfortunately, expression proposed in \cite{ZW2019} does not look useful for practical calculations since it contains the Moyal product of functions defined in phase space.
Another representation of the QHE conductivity through the two - point Green functions has been given in \cite{Imai:1990zz} (see also references therein). In principle, the idea of \cite{Imai:1990zz} is somehow similar to that of \cite{ZW2019} and of the present paper: the final form of topological expression for the Hall conductivity resembles the expression for the degree of mapping, where the mapping is defined by the Green function. In \cite{ZW2019} this expression is modified replacing the ordinary product by Moyal product, while the Green function is replaced by its Wigner transform. In \cite{Imai:1990zz} the ordinary matrix product is used. However, the Green function is taken in the specific representation, when it is represented by the infinite dimensional matrix depending on momentum. Otherwise the topological invariant of \cite{Imai:1990zz} is similar to expression of the present paper. Since the matrix is infinite dimensional, such an expression looks useless for the computational purposes. The advantage of  expression proposed by us in the present paper is that it is composed of the $N\times N$ matrices with finite $N$. Those matrices depend on momentum.

First we derive our expression for the Hall conductivity for the non - interacting model. Next, we demonstrate that it remains
valid also in the presence of interactions, when the non - interacting Green function is replaced by the complete two point Green function with the interaction corrections. These corrections are taken into account non - perturbatively through the truncated Schwinger - Dyson equation. 
It is well - known that interactions are able to lead the fermionic system in the presence of external magnetic field to the fractional QHE phases \cite{Tong:2016kpv}. We suppose that our expression might be
used for the topological description of the QHE in these phases. The fractional QHE (FQHE) may be observed when disorder is decreased. Then the additional plateaux emerge in the quantum Hall conductivity. This effect was discovered first for the fractional conductivity equal to $\nu=1/3$ of the Klitzing constant $e^2/h$ \cite{PhysRevLett.48.1559}.  It was supposed by Laughlin that the origin of the observed FQHE with $\nu = 1/3$, as well as any $n = 1/q$ with odd integer $q$ is due to the formation of the correlated incompressible electron liquid with exotic properties \cite{PhysRevLett.50.1395}. Later the theoretical description of the other types of FQHE was given including the FQHE with $\nu=2/5$ and $\nu=3/7$, as a part of the $p/(2sp\pm 1)$ series (  $s,p\in\Z$). It is widely believed that the FQHE  may be explained by the so-called {\itshape composite-fermion} theory, in which the FQHE is viewed as an integer QHE of a novel quasi-particle that consists of an electron that “captures” an even number of magnetic flux quanta \cite{PhysRevLett.63.199},\cite{PhysRevB.41.7653}. However, no precise microscopic topological explanation of the FQHE is given until now. We hope that the present paper gives a certain hint in this direction.

Notice that the AQHE existing without magnetic field can also exist in fractional form \cite{FAQHE}. The effects of interactions within the 2d topological insulators were considered
in \cite{2DTI_corr}. In graphene - like sytems relation of Coulomb interactions to the renormalization of Fermi velocity was studied, for example, in \cite{Tang2018_Science}.
Various questions related to interaction effects in 2D systems have been discussed in  \cite{corr_2d,corr_2d_2,AQHE_no_corr}.
The interaction corrections in 3D Weyl semimetals are discussed in
  \cite{corr_WSM1,corr_WSM2}.
In the present paper we consider the tight - binding models in 2d. Similar tight - binding models have been widely considered in the past (see, for example, in \cite{Z2016_1,tb2d,tb2d2,tb2d3,corr_2d} and references therein). For the description of various 3d tight - binding models see \cite{tb1,tb2,tb3,tb4,tb5}.

\section{Hall conductivity in Harper representation (non - interacting systems)}

\label{SectHarper}

In this section we start from the standard field - theoretic representation of the tight - binding model for the two - dimensional material in the presence of external magnetic field. We will introduce the notion of magnetic Brillouin zones in a manner slightly different from the standard one. It will be used latter in this form to express Hall conductivity as a topological invariant expressed through the Green functions. More details of this consideration of the non - interacting systems may be found in \cite{SZZ2020}. 

The simplest tight binding Hamiltonian (for rectangular $2D$ lattice) in the presence of electromagnetic field has the form:
\begin{eqnarray}
\^H(a^\dag,a)&=&-t\sum_x\sum_{j=1,2} \(a^\dag_x e^{-ieaA_j(x)/\hb} a_{x+e_j}+h.c.\) \nonumber\\ && + \sum_x e a^\+_x A_0(x) a_x\label{Hsimple}
\end{eqnarray}
Here sum is over the sites $x$ of rectangular $2D$ lattice, $e_j$ is the unit vector in the $j$ - th direction that connects adjacent lattice sites. Field $A_0(x)$ is responsible for external electric field $\vec{E} = - \nabla A_0(x)$. 

{\it In the following we illustrate our derivation by direct consideration of this Hamiltonian. However, the obtained expressions for the electric current and conductivity are valid also for the Hamiltonian of a more general form, which may be checked easily at each step.} For the present consideration the lattice has to be rectangular. The generalization to the lattice of general form is not described here (although such a  generalization is straightforward).

 Let us denote
\be \^Q=\sum_x a_x^+ \partial_\tau a_x +\^H(a^\+,a)-\mu \^N(a^\dag,a) =  \sum_{ij} a_i^\+ \hat{\bf Q}_{ij} a_j  \label{S000} \ee
with the fermion number operator
\be
N(a^\+,a)=\sum_x a^\+_x a_x
\ee
and derivative with respect to (imaginary) time $\tau$ (the creation and annihilation operators depend on time in Heisenberg representation).
 $\hat{\bf Q}$ will be called below Dirac operator. The partition function of the theory has the form of integral over Grassmann - valued fields $\psi_x(\tau)$
 \begin{eqnarray}
 	Z
 	&=\int D(\bar\psi,\psi)
 	e^{-\int_0^\beta d\tau
 		\bm{\bar\psi}_x(\tau) \hat{\bf Q}_{xy} \bm{\psi}_y(\tau) }
\end{eqnarray}

We consider the system in the presence of constant magnetic field, and take the electronagnetic potential in Landau gauge $A_1=0$, $A_2=Bx_1$. Green function is given as the inverse to operator $\hat{\bf Q}$:
$$
\hat{\bf G} = \hat{\bf Q}^{-1}
$$

Let us require that magnetic field is quantized and has the form
\be
\frac{2\pi}{Na}\nu=\frac{eaB}{\hb}
\ee
with mutually simple integer numbers $\nu$ and $N$.
We define magnetic flux quantum $\Phi_0=\frac{\hb}{e}2\pi$ and $\Phi=a^2B$. This gives the following condition on the magnetic flux through the lattice cell
\be
\frac{\nu}{N}=\frac{\Phi}{\Phi_0}\label{NPhi}
\ee
In practise for any value of $B$ (given by a rational number times $\frac{h}{e a^2}$) we can choose $\nu$ and $N$ to fulfill the above relation. Under these conditions the Magnetic Brillouin zones may be defined, and the Dirac operator becomes $N\times N$ matrix in the additional index. The latter representation is called Harper representation for the considered system.

Namely, we divide the Brillouin zone into the Magnetic Brillouin zones with eigenvectors of momentum
\be
\ket{(u,n),v}\equiv  \Ket{u +n\frac{2\pi}{aN}\nu,v}\label{HarperV}
\ee
Here $u \in [\Delta, \Delta + \frac{2 \pi }{a N})$ is momentum along the $x_1$ axis while $v$ is momentum along the $x_2$ axis, while $n = 0,..., N-1$. $\Delta$ may be arbitrary. 

In Matsubara representation matrix elements of Dirac operator (without the term containing external electric field)  with respect to vectors defined in Eq. (\ref{HarperV}) are  
\begin{eqnarray}
Q^{\om_m}_{uvnn'}&=&
\Big[-i \om_m-\mu-2t\cos((u,n)a)\Big]\de_{n,n'}\nonumber\\&&-t
e^{iva}\de_{n,n'-1} -t
e^{-iva} \de_{n,n'+1}
\end{eqnarray}
while
\be
(u,n)a=\(u+n\frac{2\pi}{Na}\nu\)a=ua+n\frac{2\pi}{N}\nu
\ee
i.e.
\begin{eqnarray}
Q^{\om_m}_{p_1p_2nn'}&=&
\[-i\om_m-\mu-2t\cos\(p_1a+n\frac{2\pi}{N}\nu\)\]\de_{n,n'}\nonumber\\&& -t
e^{ip_2a}\de_{n,n'-1} -t
e^{-ip_2a} \de_{n,n'+1}
\end{eqnarray}
Here $\omega_n$ are Matsubara frequencies. 
Notice that the derivation of Harper representation has been given in numerous publications (for the review see, for example, \cite{Tong:2016kpv}).

The conductivity averaged over the system area may be calculated easily through the response of electric current to electric field. Current averaged over the system area is
\be
\braket{\^{\bar J}_i} =
\frac{1}{L^2}\sum_x \braket{\^J_i(x)} =
\frac{1}{\cZ} \frac{1}{\beta L^2} \frac{\de \cZ}{\de A_i}=
\frac{1}{\beta L^2} \frac{\de \log \cZ}{\de A_i}
\ee
We assume here that the system has rectangular form with the linear size $L$.  
This expression may be represented as 
\be
\braket{\^{\bar J}_i} =
-\frac{1}{\beta L^2}\sum_{\omega_n}
\Tr\[\frac{\de\^{\bf Q}}{\de A_i} \^{\bf G} \]
\ee
where $\de A_i$ is homogeneous external electromagnetic potential. The response of electric current to constant external electric field $E$ gives conductivity. Therefore, we consider the response of Dirac operator and Green function to electric field:
\be
&\^ {\bf Q}\ra \^ {\bf Q}'=\^ {\bf Q}+\de^{\vec E} \^ {\bf Q}=\^ {\bf Q}+\frac{\partial \^ {\bf Q}}{\partial E_i}E_i \label{J000}
\ee
And the average conductivity is given by
\be
\bar \si_{ij}= \frac{1}{\beta L^2}\sum_{\omega_n}
\Tr\[\frac{\partial \^{{\bf Q}}}{\partial  A_i}\^ {\bf G} \frac{\partial  \^ {\bf Q}}{\partial  E_j} \^ {\bf G} \]
\ee
Next, we come to Matsubara representation and insert into this expression the completeness relation that introduces the magnetic Brillouin zone (see also Appendix \ref{AppD}):
\begin{eqnarray}
1&=&\sum_{n=0}^{N-1}
\int_{-\infty}^{\infty} du
\int_{0}^{\frac{2\pi}{a}} dv
\Ket{u +n\frac{2\pi}{aN}\nu,v}\nonumber\\&& \Bra{u +n\frac{2\pi}{aN}\nu,v}
\te\(u-\De\)\te\(\De+\frac{2\pi}{Na}-u\)\label{compl}
\end{eqnarray}
As a result we obtain the expression for conductivity through $\bf Q $ and $\bf G$ written in Harper representation. 

At low temperature $T\ra 0$ the sum over Matsubara frequencies may be replaced by an integral $\sum_{\om_m}\ra \frac{1}{2\pi T}\int d\om$
\be
\bar \si_{ij}&=
-\frac{e^2}{2\pi \hb  4\pi^2 N}
\int d\om \intl_{BZ} dp_1 \intl_{BZ} dp_2 \\& \Tr
\[
\frac{\pd {\bm Q}_{p_1p_2}^\om}{\pd p_i}
{\bm G}_{p_1p_2}^\om
\frac{\pd {\bm Q}_{p_1p_2}^\om}{\pd \om}
{\bm G}_{p_1p_2}^\om
\frac{\pd {\bm Q}_{p_1p_2}^\om}{\pd p_j}
{\bm G}_{p_1p_2}^\om
\]=\\
&=-\frac{e^2}{2\pi \hb 4 \pi^2 N }
\int d\om \intl_{BZ} dp_1 \intl_{BZ} dp_2
\\&\sum_{a,...,f=1}^N
\[
\frac{\pd Q_{p_1p_2ab}^\om}{\pd p_i}
G_{p_1p_2bc}^\om
\frac{\pd Q_{p_1p_2cd}^\om}{\pd \om}
G_{p_1p_2de}^\om
\frac{\pd Q_{p_1p_2ef}^\om}{\pd p_j}
G_{p_1p_2fa}^\om
\]\label{sfin}
\ee
Notice that although Dirac operator $\bf Q$ is taken here in Harper representation the integrals in the above expression are within the whole Brillouin zone. Here by  ${\bm Q}_{p_1p_2}^\om$ we denote $N\times N$ matrix with components $Q_{p_1p_2ab}^\om$ ($a,b = 1,..., N$).

For the antisymmetric (Hall) part of conductivity we obtain:
\begin{eqnarray}
\bar \si^{AS}_{ij}&=&
\epsilon_{ij}\, \frac{e^2}{h} \, \frac{1}{N}\, \frac{\epsilon^{abc}}{24  \pi^2} \,
\int d\om \intl_{BZ} dp_1  dp_2 \label{HC} \\&& \Tr
\Bigl[
\frac{\pd {\bm Q}_{p_1p_2}^\om}{\pd p_a}
{\bm G}_{p_1p_2}^\om
\frac{\pd {\bm Q}_{p_1p_2}^\om}{\pd p_b}
{\bm G}_{p_1p_2}^\om
\frac{\pd {\bm Q}_{p_1p_2}^\om}{\pd p_c}
{\bm G}_{p_1p_2}^\om
\Bigr] \nonumber
\end{eqnarray}

{This is the main result of our paper. It is valid for the non - interacting tight - binding fermionic system of general form although the derivation has been illustrated by consideration of the model with the Hamiltonian of Eq. (\ref{Hsimple}).} In Eq. (\ref{HC}) integral is over the whole Brillouin zone where matrix elements of $\bf Q$ obey periodic boundary conditions. As a result the given expression is topological. It has been derived for the system defined on rectangular lattice with the magnetic flux through the elementary lattice cell given by a rational number $\nu/N$ times elementary magnetic flux. One can see that the Hall conductivity may be represented in the form of the product
 $
{\sigma_H = \frac{e^2}{h} \, \frac{1}{N}\,{\cal N}}
$
where $\cal N$ is the integer topological invariant composed of the $N\times N$ matrices $\bf Q$.  For the non - interacting systems this invariant {\it should} be equal to the integer multiple of $N$ in order to provide the integer QHE.

We expect that Eq. (\ref{HC}) remains valid also for the case of an interacting system, where matrix $\bf G$ is replaced by the complete propagator defined in the magnetic Brillouin zone. In this case the value of $\cal N$ is not necessarily equal to the integer multiple of $N$. This way we may obtain the topological description of the fractional quantum Hall effect.

\section{Numerical results in non - interacting systems}
\label{SectionIV}
\subsection{ Diophantine equation.}

In this section we illustrate the general expressions obtained above by numerical results obtained for the system with the one - particle Hamiltonian of the form
\be
\^H(a^\dag,a)&=v_1\sum_x \(a^\dag_x e^{-ieaA_1(x)/\hb} a_{x+e_1}+h.c.\) \\& + v_2\sum_x \(a^\dag_x e^{-ieaA_2(x)/\hb} a_{x+e_2}+h.c.\)\\&+  \sum_x e a^\+_x A_0(x) a_x
\ee
with the two different hopping parameters $v_1$ and $v_2$.

We consider the case of constant magnetic field originated from  potential
$$
A_1(x_1,x_2) = B x_2, \quad A_2(x_1,x_2) = 0
$$
The values of magnetic flux through the lattice cell are chosen to be equal to $1/N$ of elementary flux $\Phi_0 = h/e$. The two possibilities are considered: with $N = 3$ and $N=4$. We calculate numerically spectrum of the system, and the value of Hall conductivity using the obtained above expression
\be
\bar \si^{AS}_{ij}&=
\epsilon_{ij}\, \frac{e^2}{h} \, \frac{1}{N}\, \frac{\epsilon^{abc}}{24  \pi^2} \,
\int d\om \intl_{BZ} dp_1  dp_2\\& \Tr
\[
\frac{\pd {\bm Q}_{p_1p_2}^\om}{\pd p_a}
{\bm G}_{p_1p_2}^\om
\frac{\pd {\bm Q}_{p_1p_2}^\om}{\pd p_b}
{\bm G}_{p_1p_2}^\om
\frac{\pd {\bm Q}_{p_1p_2}^\om}{\pd p_c}
{\bm G}_{p_1p_2}^\om
\]\label{sHall}
\ee

In both cases we reproduce the known result for spectrum (that results from the solution of Harper equation). Moreover, in the case when the chemical potential belongs to the gap, we reproduce the value of Hall conductivity that might me obtained alternatively using the solution of Diophantine equation. This confirmes the validity of the derived general expression for the Hall conductivity as an integral in the Brillouin zone.

Recall that the standard method for calculation of Hall conductivity gives
$$\sigma_H = \frac{e^2}{h}\,t_r$$
 where $t_r$ is the solution of Diophantine equation
 $$
 r = N s_r + \nu t_r
 $$
for integer $r,s_r,t_r$. In our case $N = 3$ or $4$ while $\nu = 1$. Here $|t_r| \le N/2$, while $r = 1,..., N$.

One can check that both for $N=3$ and $N=4$ there are
two nontrivial solutions of this equation with $t_r = \pm 1$. This pattern is reproduced by our numerical results and is illustrated by Fig. \ref{fig.topoNum_3} and Fig. \ref{fig.topoNum_4}.

\subsection{The case of N=3}

In this section, we consider the topological number $\sigma_H$ defined above, in the
presence of magnetic field $\cal B$, which satisfies ${\cal B}a^2=\phi_0/3$.
Here, $\phi_0$ is the magnetic flux quantum,
$\phi_0=h/e$. For simplicity we use here unities with $\hbar = e = 1$. In these units the value of conductivity is quantized as an integer multiple of $1/(2\pi)$.

If one switches on a small electric field $\cal E$ along $x$-axis, according to Eq. (\ref{J000})
the Hall current density along $y$-axis $J_2$ is given by
\begin {eqnarray}\label{Hall_current_density}
J_2=\int_{-\infty}^{+\infty}\frac{dE}{2\pi}
\int_{-\pi/3}^{+\pi/3}\frac{d p_1}{2\pi}
\int_{-\pi}^{+\pi}\frac{d p_2}{2\pi}\, Tr\,
G(p) \frac{\partial Q}{\partial p_2}
\end{eqnarray}
where $Q$ is the matrix
\begin {eqnarray}\label{Q_matrix}
\begin{pmatrix}
E - 2 v_1 cos(p_1)  & -v_2 e^{ip_2}         & -v_2 e^{-ip_2}      \\
-v_2 e^{-ip_2}   & E - 2 v_1 cos(p_1+2\pi/3) &  -v_2 e^{ip_2}     \\
-v_2 e^{ip_2}    & -v_2 e^{-ip_2}    & E -2 v_1 cos(p_1-2\pi/3)
\end{pmatrix}\nonumber
\end{eqnarray}
In the leading order, $G=G^{(1)}(p)$ which is given by
\begin {eqnarray}\label{G1}
G^{(1)}(p)=\frac{i}{2} G^{(0)}\frac{\partial Q}{\partial p_i}
\frac{\partial G^{(0)}}{\partial p_j} F_{ij},
\end{eqnarray}
with $ G^{(0)}=1/Q$ and $i,j\in \{0,1\}$. Therefore,
$G^{(1)}=(i/2) G^{(0)}(\partial_1 Q \partial_0 G^{(0)}
-\partial_0 Q \partial_1 G^{(0)})(i \cal E)$, and
\begin {eqnarray}\label{Hall_current_density_2}
J_2&=&\frac{-{\cal E}}{2}
\int_{-\infty}^{+\infty}\frac{dE}{2\pi}
\int_{-\pi/3}^{+\pi/3}\frac{d p_1}{2\pi}
\int_{-\pi}^{+\pi}\frac{d p_2}{2\pi}\nonumber\\&&
Tr G^{(0)}(\frac{\partial Q}{\partial p_1}\frac{\partial G^{(0)}}{\partial p_0}
-\frac{\partial Q}{\partial p_0}\frac{\partial G^{(0)}}{\partial p_1})
\frac{\partial Q}{\partial p_2}.
\end{eqnarray}

With the help of Mathematica, we obtained
\begin {eqnarray}\label{trace}
&& {\rm Tr} \Big( G^{(0)}\big(\frac{\partial Q}{\partial p_1}\frac{\partial G^{(0)}}{\partial p_0}-
\frac{\partial Q}{\partial p_0}\frac{\partial G^{(0)}}{\partial p_1}\big)
\frac{\partial Q}{\partial p_2} \Big)  \nonumber\\
&=& 6\sqrt{3} u^2 i \, \frac{E-U}{E^3-\frac{12+3u^3}{4}E -V},
\end{eqnarray}
where $U=\frac{u}{4}cos(3p_1)+\frac{1}{2}cos(3p_2)$
and $V=\frac{u^3}{4}cos(3p_1)+2 cos(3p_2)$, with $u=2v_1/v_2$.
Note that if one wants to take chemical potential into account, one will
change $E$ into $E+\mu$. From the denominator of Eq.(\ref{trace}), we
can find the energy spectra of the system. The energy levels will be
the solutions of the cubic equation
$4(E/2\alpha)^3 -3(E/2\alpha)=V/2\alpha^3$,
with $\alpha=\sqrt{u^2+4}/2$.
Explicitly, the solutions (from small to big) are given by
\begin {eqnarray}\label{roots}
E_1&=& 2\alpha \, cos(\frac{\theta+2\pi}{3})  \nonumber\\
E_2&=& 2\alpha \, cos(\frac{\theta-2\pi}{3})  \nonumber\\
E_3&=& 2\alpha \, cos(\frac{\theta}{3}),
\end{eqnarray}
with $\theta= arccos(V/2\alpha^3)\in [0,\pi]$.

\begin{figure}[h]
\centering  %
\includegraphics[width=10cm]{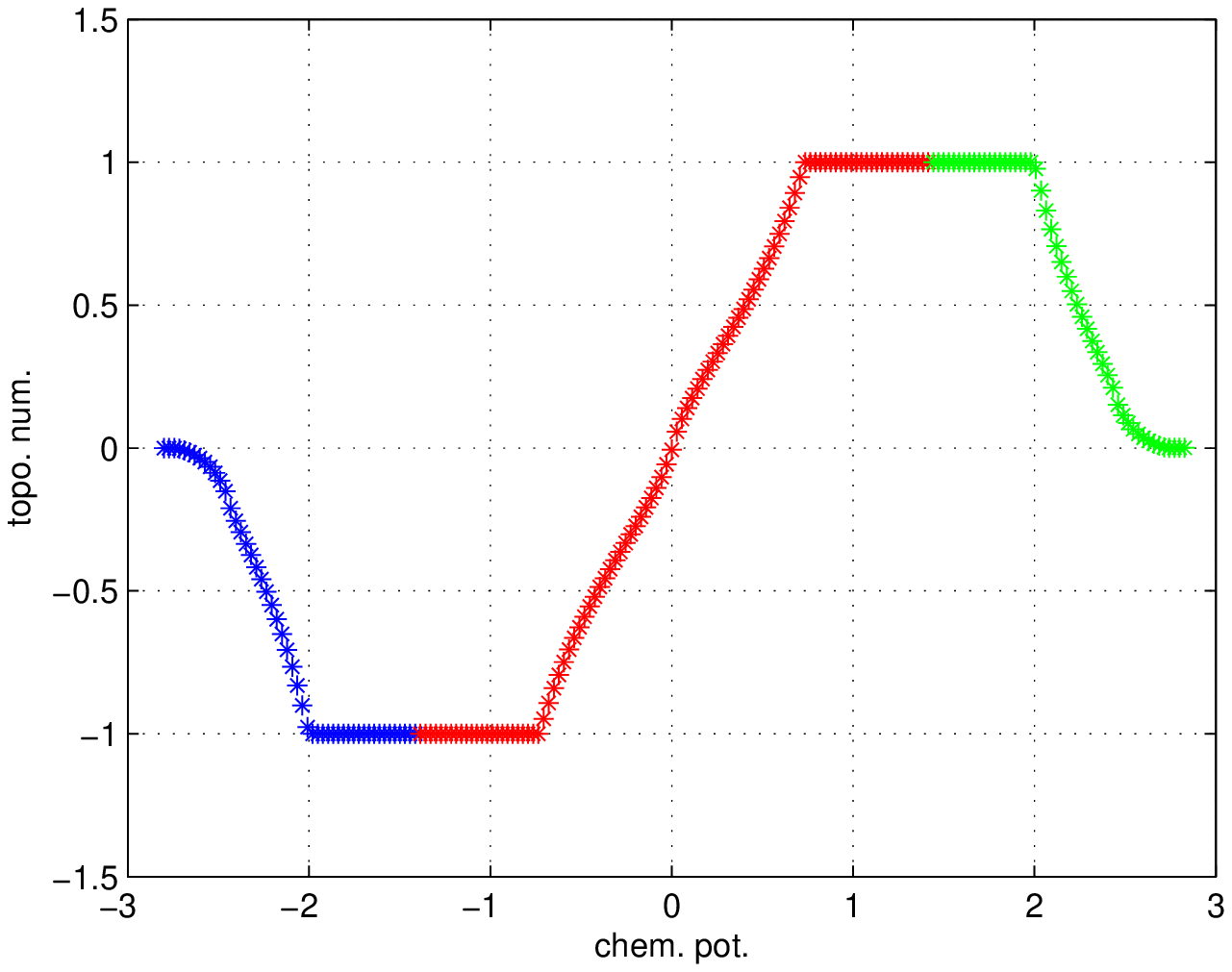} 
\caption{Topological number v.s. chemical potential (expressed in lattice units), for the case of $N=3$ and $u=2$.}  %
\label{fig.topoNum_3}   %
\end{figure}

Next step is to calculate the Hall conductivity from
Eq.(\ref{Hall_current_density}). Our method is to integrate
$dE$ analytically and then compute the integral $d^2 {\bf p}$
numerically. Taking the case of $\mu< inf\{E_2\} $,
as an example, we replace $E$ by $E+\mu+i\delta_{\bf p}$
with $\delta_{\bf p}=\eta\, \Theta(E_1(p)-\mu)$.
Applying residue theorem to the integration of dE,
we obtained
\begin {eqnarray}\label{Hall_current_density_3}
J_2&=&{\sqrt{3}\cal E}
\int_{-\pi}^{+\pi}\frac{d a}{2\pi}
\int_{-\pi}^{+\pi}\frac{d b}{2\pi}
(E_1-E_2)^{-2}(E_3-E_1)^{-2}\nonumber\\&&[1-2(U-E_1)(\frac{1}{E_1-E_2}-\frac{1}{E_3-E_1})],
\end{eqnarray}
where $E_i$'s are the roots, but as the functions of variables $a=3p_1$ and $b=3p_2$.
Then using Matlab, we obtained the numerical results shown in Fig.\ref{fig.topoNum_3},
via numerical integration.


%
\begin{figure}[h]
\centering  %
\includegraphics[width=10cm]{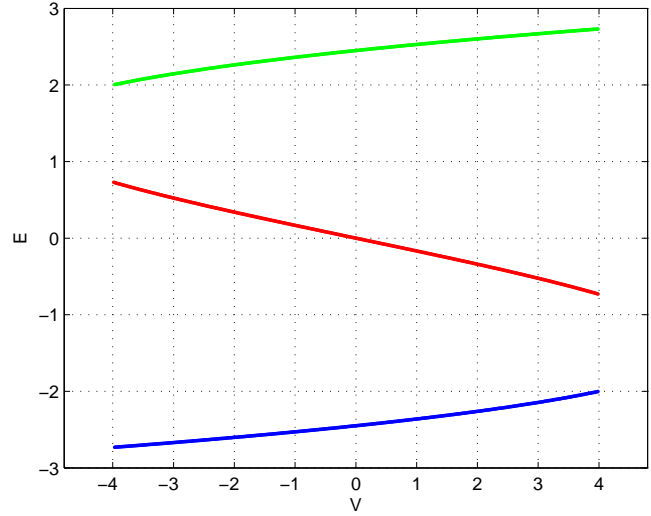} 
\caption{Energy spectra of N=3, with $u=2$. In the x-axis: $x=V$.}  %
\label{fig.spectra_3}   %
\end{figure}

\subsection{The case of N=4}

In this section, we consider the case of $N=4$, i.e.
the magnetic field $\cal B$ satisfies ${\cal B}a^2=\phi_0/4$,
with $\phi_0=h/e$.
If one switches on a small electric field $\cal E$ along $x$-axis,
the Hall current density along $y$-axis $J_2$ is given by
\begin {eqnarray}\label{Hall_current_density_q=4}
J_2=\int_{-\infty}^{+\infty}\frac{dE}{2\pi}
\int_{-\pi/4}^{+\pi/4}\frac{d p_1}{2\pi}
\int_{-\pi}^{+\pi}\frac{d p_2}{2\pi}\, Tr\, 
G(p) \frac{\partial Q}{\partial p_2}
\end{eqnarray}
where $Q$ is the matrix
\begin {eqnarray}\label{Q_matrix}
\begin{pmatrix}
Q_{11} & Q_{12}\\ Q_{21} & Q_{22}
\end{pmatrix}
\end{eqnarray}
with
\begin {eqnarray}
Q_{11} = 
\begin{pmatrix}
E - 2 v_1 cos(p_1)  & -v_2 e^{ip_2}           \\
-v_2 e^{-ip_2}      & E - 2 v_1 cos(p_1+\pi/2)
\end{pmatrix}
\end{eqnarray}
\begin {eqnarray}
Q_{12} = 
\begin{pmatrix}
0                      & -v_2 e^{-ip_2}      \\
  -v_2 e^{ip_2}       & 0  
\end{pmatrix}
\end{eqnarray}
\begin {eqnarray}
Q_{21} = 
\begin{pmatrix}   
0                   & -v_2 e^{-ip_2}        \\
-v_2 e^{ip_2}      & 0  
\end{pmatrix}
\end{eqnarray}
\begin {eqnarray}
Q_{22} = 
\begin{pmatrix} 
  E -2 v_1 cos(p_1+\pi)  &   -v_2 e^{ip_2} \\
 -v_2 e^{-ip_2}          & E -2 v_1 cos(p_1-\pi/2)
\end{pmatrix}
\end{eqnarray}

Using Mathematica, we obtained
\begin {eqnarray}\label{trace_q=4}
&& {\rm Tr} \Big( G^{(0)}\big(\frac{\partial Q}{\partial p_1}\frac{\partial G^{(0)}}{\partial p_0}-
\frac{\partial Q}{\partial p_0}\frac{\partial G^{(0)}}{\partial p_1}\big)
\frac{\partial Q}{\partial p_2} \Big)  \nonumber\\
&=& 16 u^2 i \, \frac{E(E^2+U)}{[E^4-(u^2+4)E^2 + V]^2},
\end{eqnarray}
where $U=\frac{u^2}{4}(1-cos(4p_1))+(1-cos(4p_2))$
and $V=\frac{u^4}{8}(1-cos(4p_1))+2 (1-cos(4p_2))$, with $u=2v_1/v_2$.
Note that if one wants to take chemical potential into account, one will
change $E$ into $E+\mu$. From the denominator of Eq.(\ref{trace_q=4}), we
can find the energy spectra of the system. The energy levels will be
the solutions of the quadratic equation
$E^4 -(u^2+4)E^2+V=0$.
Explicitly, the solutions (from small to big) are given by
\begin {eqnarray}\label{roots_q=4}
E_1&=& -(w+\sqrt{w^2-V})^{1/2}  \nonumber\\
E_2&=& -(w-\sqrt{w^2-V})^{1/2}  \nonumber\\
E_3&=& (w-\sqrt{w^2-V})^{1/2}  \nonumber\\
E_4&=& (w+\sqrt{w^2-V})^{1/2}  \nonumber\\
\end{eqnarray}
with $w=u^2/2+2$.

\begin{figure}[h]
\centering  %
\includegraphics[width=10cm]{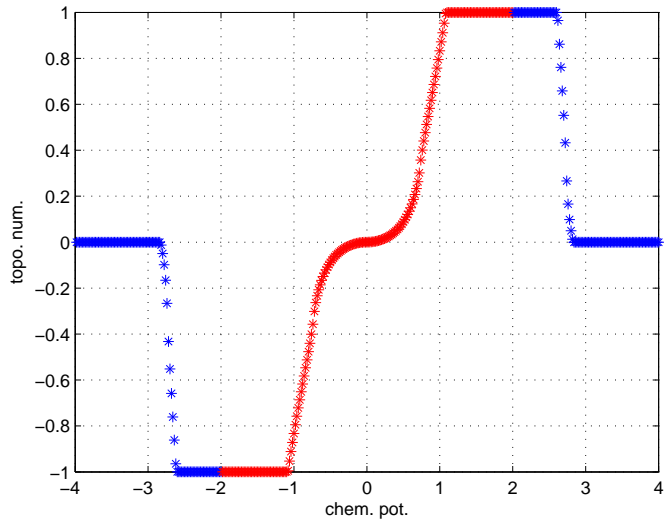} 
\caption{Topological number v.s. chemical potential (expressed in lattice units), for the case of $N=4$ and $u=2$.}  %
\label{fig.topoNum_4}   %
\end{figure}

Next step is to calculate the Hall conductivity from
Eq.(\ref{Hall_current_density_q=4}). Our method is to integrate
$dE$ analytically and then compute the integral $d^2 {\bf p}$
numerically. Taking the case of $\mu< inf\{E_2\} $,
as an example, we replace $E$ by $E+\mu+i\delta_{\bf p}$
with $\delta_{\bf p}=\eta\, \Theta(E_1(p)-\mu)$.
Applying residue theorem to the integration of dE,
we obtained
\begin {eqnarray}\label{Hall_current_density_q=4_b}
J_2&=&{2u^2\cal E}
\int_{-\pi}^{+\pi}\frac{d a}{2\pi}
\int_{-\pi}^{+\pi}\frac{d b}{2\pi}
(E_1-E_2)^{-2}(E_1-E_3)^{-2}\nonumber\\&&(E_1-E_4)^{-2}
[3E_1^2+U-2(E_1^3+UE_1)H],
\end{eqnarray}
where $H=\frac{1}{E_1-E_2}+\frac{1}{E_1-E_3}+\frac{1}{E_1-E_4}$,
and $E_i$'s are the roots, but as the functions of variables $a=4p_1$ and $b=4p_2$.
Then using Matlab, we obtained the numerical results shown in Fig.\ref{fig.topoNum_4} ,
via numerical integration.


%
\begin{figure}[h]
\centering  %
\includegraphics[width=10cm]{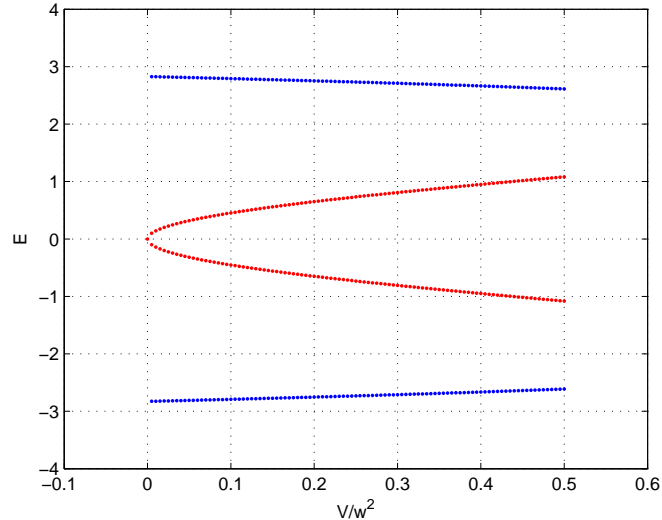} 
\caption{Energy spectra of N=4, with $x=V/w^2$, and $u=2$.}  %
\label{fig.spectra_4}   %
\end{figure}

\section{Quantum hall effect in (2+1)D in the presence of interactions. Perturbation theory.}

Our consideration proposed in the present section is based essentially on those of \cite{ZZ2021}. We will see that the  expression for Hall conductivity through the Green functions remains valid in the presence of interactions if those interactions are taken into account within perturbation theory. (Non - interacting Green function is to be substituted in the non - interacting expression instead of the non - interacting Green function.) It appears, however, that since the value of Hall conductivity   is a topological invariant, its value cannot be changed by interactions taken into account perturbatively when coupling constant grows smoothly from zero to a given value, unless the phase transition is encountered. Indeed,  the perturbation corrections taken into account order by order cannot modify the value of Hall conductivity.  

We consider a finite sized, periodic, lattice regularized fermionic system in (2+1)D in the presence of interactions. Formally the system may be regarded as being defined on a 2D-torus in the spatial directions. Finite periodicity in imaginary time corresponds to a system at finite temperature. Infinite size limits may later be taken both for the spatial and the imaginary time direction. The latter limit corresponds to a transition to zero temperature. The theories under consideration include exchange by scalar bosons. The particular cases are: Yukawa interaction, Coulomb interaction as well as a fermion four-point interaction. The most important case here is the Coulomb interactions, which give the dominant contribution to formation of FQHE.

The action of the free fermion system is taken to be of the form

\begin{align}
	S_0=\int d\tau \sum_{x^{\prime},x}\bar{\psi}_{x^{\prime}}(\tau )Q^{\psi}_{x^{\prime},x}\psi_x(\tau ),
\end{align}
Operator $Q$ here is of a general form. For concreteness we give below the particular   form that corresponds to an idealized system defined on the rectangular lattice (this systems differs somehow from the one of Eq. (\ref{S000})). {\it Anyway, as in the above sections, here the results will not depend on the particular form of $Q$.}

The part of the action comprising the interacting theories may have, for example, the following forms:

\begin{align}
	S_{\eta}=&S_0+\int d\tau \sum_{x,x^{\prime}}\phi_{x^{\prime}}(\tau )Q^{\phi}_{x^{\prime},x}\phi_x(\tau )-\eta \sum_x\bar{\psi}_x(\tau )\psi_x (\tau )\phi_x(\tau ),\\
	S_{\alpha}=&S_0-\alpha\int d\tau \sum_{x^{\prime},x}\bar{\psi}_{x^{\prime}}(\tau )\psi_{x^{\prime}}(\tau )V(x^{\prime}-x)\bar{\psi}_x(\tau )\psi_x(\tau ),\\
	S_{\lambda}=&S_0-\lambda\int d\tau \sum_{x^{\prime},x}\bar{\psi}_{x^{\prime}}(\tau )\psi_{x^{\prime}}(\tau )W(x^{\prime}-x)\bar{\psi}_x(\tau )\psi_x(\tau )
\end{align}
which correspond respectively to exchange by a scalar boson field $\phi$ (in this case we also need to add the action of this field itself - the term containing operator $Q^\phi$), and the particular forms of this interaction: Coulomb interactions and contact four - fermion interactions. 

Here are particular examples of the fermion action and the scalar boson action: 

\begin{align}
	&Q^{\psi}_{x^{\prime},x}=i(i\partial_{\tau}-(A_3)_x(-i\tau )\delta_{x^{\prime},x}-\frac{1}{2}\sum_{i=1,2}[(1+\sigma^i)\delta_{x^{\prime},x+e_i}e^{iA_{x+e_i,x}}+(1-\sigma^i)\delta_{x^{\prime},x-e_i}e^{iA_{x-e_i,x}}]\sigma^3\\
	&\,\,\,\,\,\,\,\,\,\,\,\,\,\,\,\,\,\,\,\,+i(m+2)\delta_{x^{\prime},x}\sigma^3,\\
	&Q^{\phi}_{x^{\prime},x}=\partial^2_{\tau}\delta_{x^{\prime},x}+\sum_{i=1,2}(\delta_{x^{\prime},x+e_i}+\delta_{x^{\prime},x-e_i}-(M^2+4)\delta_{x^{\prime},x}),\\
	&V(x^{\prime}-x)=\frac{1}{|x^{\prime}-x|},\,\,\,\, W(x^{\prime}-x)=1
\end{align}

The parameters $m$ and $M$ represent the fermion and boson masses, respectively. Furthermore we employ the notation $A_{u,v}=\int_v^uA\cdot ds$ where the integral is along the straight line connecting $u$ and $v$. In this section we consider field $A$ depending on continuous coordinates because we are going to use Wigner - Weyl calculus. The expressions represent the respective action at finite temperature with imaginary time $\tau =it\in [-\frac{\beta}{2},\frac{\beta}{2}]$ and imaginary time periodicity of $\beta =\frac{1}{T}$ for temperature $T$. The respective expressions at zero temperature are obtained by replacing $\tau $ with $t$ and taking the limit $T\to 0$.

The fermions are also coupled to a classical external source given by an electromagnetic gauge field $A_{\mu}$ which is included already in $S_0$. This electromagnetic field comprises a constant external electric field $E$, a constant external magnetic field $B$ orthogonal to the spatial directions. The respective contributions will be split by $A^{E=0}$ and $A^E$, where $E$ denotes the constant external electric field, respectively, while their sum is represented by $A$. We will assume here that spatial inhomogeneities are insignificant on the scale of the lattice spacing and that sums over lattice sites may be well approximated by spatial integrals. We further work in lattice units assuming a rectangular lattice structure. For a rectangular lattice with lattice spacing $a$ we therefore set $a=1$. Furthermore we use the systm of units with $k=\hbar =1$, where  $k$ is the Boltzmann constant and $\hbar$ the reduced Planck constant. Electric charge of electron is included into the definition of the electromagnetic field.

Let us consider the case of spatial homogeneity first. The electromagnetic current in the free theory is given by the variation of the effective action with respect to the external electromagnetic gauge field. Under the assumption that the inverse propagator $Q^{\psi}$ is a function of the combination $p-A(x)$ only  it follows that 

\begin{align}
	J^k_0(x)=\frac{\delta log(Z)}{\delta A_k(x)}=-\int  {d^3p_1}{d^3p_2} Tr((G_0)(p_1,p_2)\frac{\partial (Q_0)(p_2,p_1)}{\partial p_k})
	\label{homocurrent}
\end{align}

at zero temperature with partition function

\begin{align}
	Z=\int D(\bar{\psi},\psi )e^{-S_0(\bar{\psi},\psi )}.
\end{align}

At finite temperature the integral over the time component of momentum space simply becomes a sum over Matsubara modes divided by $\beta =\frac{1}{T}$ with temperature $T$. We employ the notation 

\begin{align}
	&\Phi_{x^{\prime},x}=i(i\partial_{\tau}-(A_3)_x(-i\tau ))\delta_{x^{\prime},x},\,\,\,\, (T^i_+)_{x^{\prime},x}=-\frac{1}{2}(1+\sigma^i)\delta_{x^{\prime},x+e_i}e^{iA_{x+e_i,x}}\sigma^3=(L^i_+)_{x^{\prime},x}e^{iA_{x+e_i,x}},\\
	& (T^i_-)_{x^{\prime},x}=-\frac{1}{2}(1-\sigma^i)\delta_{x^{\prime},x-e_i}e^{iA_{x-e_i,x}}\sigma^3=(L^i_-)_{x^{\prime},x}e^{iA_{x-e_i,x}}.
\end{align}

in order to decompose the inverse propagator $Q^{\psi}_{x^{\prime},x}$. In Landau gauge we have $A_1(x)=0$ and $A_2(x)=Bx_1$, while $A_3(x)=-Ex_l$ with $l=1,2$. The associated Fourier transformed quantities $\tilde{T}^i_{\pm}(p^{\prime},p)$ and $\tilde{L}^i_{\pm}(p^{\prime},p)$ defined by

\begin{align}
	&(T^i_{\pm})_{x^{\prime},x}=\int \frac{d^2p^{\prime}}{\sqrt{(2\pi )^2}}\frac{d^2p}{\sqrt{(2\pi )^2}}e^{ip^{\prime}x^{\prime}}\tilde{T}^i_{\pm}(p^{\prime},p)e^{-ipx}\\
	&(L^i_{\pm})_{x^{\prime},x}=\int \frac{d^2p^{\prime}}{\sqrt{(2\pi )^2}}\frac{d^2p}{\sqrt{(2\pi )^2}}e^{ip^{\prime}x^{\prime}}\tilde{L}^i_{\pm}(p^{\prime},p)e^{-ipx}
\end{align}

are related by 

\begin{align}
	\tilde{T}^i_{\pm}(p^{\prime}_1,p^{\prime}_2,p_1,p_2)=\tilde{L}^i_{\pm}(p^{\prime}_1,p^{\prime}_2,p_1\pm B\delta_{i,2},p_2)
\end{align}

in Landau gauge.\\
It can be observed that the magnetic field induces a relative shift in momentum space between the Fourier transforms of $T^i_{\pm}$ and $L^i_{\pm}$.

We now treat the more general case with spatial inhomogeneity in detail first before reducing it to the homogeneous case in Harper representation.

In the presence of spatial inhomogeneities the free theory electric current may be expressed by the use of the Wigner transform as 

\begin{align}
	J^k_0(x)=-\int \frac{d^3p}{(2\pi )^3}Tr((G_0)_W(x,p)\frac{\partial (Q_0)_W(x,p)}{\partial p_k})
	\label{inhomocurrent}
\end{align}

where the subscript $0$ stands for free theory, while the subscript $W$ marks the Wigner transform (for its detailed description see \cite{ZZ2021}). Notice, that here we use the so - called approximate version of Wigner - Weyl calculus, which is valid in the presence of weak inhomogeneities.  The definitions of Wigner transformation (Weyl symbol) are given in Appendix \ref{AppC} for the case of continuous systems. These definitions are used without modifications for the lattice systems when external fields vay slowly at the distance of the order of lattice spacing. Then the sum over the lattice points may everywhere be replaced by integrals. In practise this approximation works perfectly if magnetic field strength is much smaller than $10^5$ Tesla, which takes place in all realistic solid state systems.  

 The changes of subscript $0\mapsto \{ \eta ,\alpha ,\lambda \}$ in the current and propagator (but not inverse propagator!) label the respective interacting theory expressions. Under the assumptions of constant external field strength (meaning constant electric and magnetic field) as well as functional dependencies on $p-A(x)$ only the periodic boundary conditions in momentum space due to lattice regularization imply the identity

\begin{align}
	J^k_0(x)=&-\int \frac{d^3p}{(2\pi )^3}Tr((G_0)_W(x,p)\frac{\partial (Q_0)_W(x,p)}{\partial p_k})\\
	&-\int \frac{d^3p}{(2\pi )^3}Tr((G_0)_W(x,p)\star\frac{\partial (Q_0)_W(x,p)}{\partial p_k})
\end{align}

at zero temperature. Symbol $\star$ represents the Moyal star product
$$
\star = {\rm exp} \Big(\frac{i}{2} \left( \overleftarrow{\partial}_{x}\overrightarrow{\partial_p}-\overleftarrow{\partial_p}\overrightarrow{\partial}_{x}\right) \Big)
$$

 The same conclusion carries over to the interacting systems. If the Wigner transformed inverse propagator depends only on the difference $p-A(x)$, it can be shown by induction starting from $(G_0)_W\star (Q_0)_W=1$ that the same follows for $(G_0)_W$. The propagator will also be a polynomial in both electric and magnetic fields from which we discard all but the terms linear in the electric field and assume small electric field in the discussion of the quantum Hall effect.

As soon as the electromagnetic field is allowed to vary spatially, the insertion of the Moyal star product in place of ordinary multiplication is only allowed for the consideration of the space and time averaged current (where we introduce imaginary time extent $\beta$ and periodicity in time)

\begin{align}
	\bar{J}^k_0=\frac{1}{\beta L^2}\int d^3xJ^k_0(x).
	\label{inhomoaveragecurrent}
\end{align}

in the absence of constant electric field. How to include it will be discussed below. The parameter $L$ represents the spatial extension of the torus in each direction. This also carries over to the interacting theories.

At finite temperature and in thermal equilibrium the Wigner transformations of the Dirac operator as well as the (interacting) Green function do not depend explicitly on imaginary time. The momentum space integral for the time component becomes a sum over discrete Matsubara modes. The Moyal star product is then only acting on spatial coordinates and their respective momenta. The previous considerations for the case of zero temperature carry over identically, except for the case of constant external electric field whose presence is essential for the quantum Hall effect. The continuous functional dependence on $p-A(x)$ of the involved functions can no longer be used for the time component to insert the Moyal star product in place of ordinary multiplication, as there appears no longer an integral over the momentum variable corresponding to the temporal component. A spatial averaging does also not amend this problem, as spatial periodicity is not fulfillable in the presence of a constant electric field. This has not been an issue at zero temperature for constant field strength, however. At both zero and nonzero temperature, the electric field explicitly breaks the periodicity assumption which requires resolution.

A solution may be provided by the following Gedankenexperiment. Divide the torus into two identical parts whose boundary is orthogonal to the external constant electric field which we assume to point in one of the two torus directions (there is no rotational invariance). Then switch the direction of the electric field within one half. This restores spatial periodicity in the direction of the electromagnetic field. It is important to note that we will assume that fermions in both halves are coupled equally to the external electric field, while their mutual interactions may be different in the two halves. More precisely we will shortly assume that the mutual interactions vanish in one half, while they are present in the other. In this way it will be possible to conclude the non-renormalization of the electric conductivity in the quantum Hall effect from that of the electric current.

Under the mentioned circumstances (including the setup presented within the Gedankenexperiment) the replacement of ordinary multiplication with the star product and vice versa is allowed anywhere in the integrated expressions. The periodicities in momentum and coordinate space are conserved under multiplication with the Moyal star product. We may therefore make use of the results of \cite{ZZ2021} on the non-renormalization of the electric current.

Starting from the expression of the current in Eq. (\ref{inhomocurrent}) we employ expansion in the assumed small contribution $F^E$ within the identity

\begin{align}
	(G_0)_W(x,p)\star (Q_0)_W(p-A(x))=1.
	\label{identityfree}
\end{align}

We write $(G_0)_W=\sum_{n=0}^{\infty}(G_0)^n_W$ with $(G_0)^n_W\propto (F^E)^n$ and keep only terms of the orders $n=0,1$. This leads to

\begin{align}
	&(G_0)^0_W(x,p)\star (Q_0)_W(p-A^{}(x))=1,
\end{align}
and 
\begin{align}
	(G_0)^1_W(x,p)\approx 
	&-\frac{i}{2}(G_0)^0_W(x,p)\star \frac{\partial (Q_0)_W(p-A^{}(x))}{\partial p_{\mu}}\star \frac{\partial (G_0)^0_W(x,p)}{\partial p_{\nu}}F^E_{\mu\nu}.
\end{align}

with $F^E_{\mu\nu}=\partial_{\mu}A^E_{\nu}(x,p)-\partial_{\nu}A^E_{\mu}(x,p)$. The contribution to the current proportional to electric field strength may then be expressed within this approximation as

\begin{align}
	J^k_0(x)\approx\frac{i}{2}\int \frac{d^3p}{(2\pi )^3}Tr((G_0)^0_W(x,p)\star \frac{\partial (Q_0)_W(p-A^{}(x))}{\partial p_{\mu}}\star \frac{\partial (G_0)^0_W(x,p)}{\partial p_{\nu}}\frac{\partial (Q_0)_W(p-A^{}(x))}{\partial p_k})F^E_{\mu\nu}.
	\label{approxinhomocurrent}
\end{align}

The only nonzero components of the field strength are $F^E_{l3}=-F^E_{3l}=iE_l$ for an electric field in direction $l$ with $l=1,2$. The ordinary product in (\ref{approxinhomocurrent}) may be replaced by a Moyal star product for the averaged current (\ref{inhomoaveragecurrent}), as the electric field is no longer contained within the propagator as well as the inverse propagator. In linear response theory we can omit in $A(x)$ the contribution of electric field, and define $W_m(x,p)=(G_0)^0_W(x,p)\star \partial_{p_m}(Q_0)_W(p-A^{E=0}(x))$ it follows that

\begin{align}
	\bar{J}^k_0=\epsilon_{ijk}\frac{\sigma_0}{2i}F^E_{ij}=\epsilon_{kl}\sigma_0E_l,\,\,\,\, \sigma_0=\frac{1}{3!}\epsilon_{abc}\int\frac{d^3x}{\beta L^2}\frac{d^3p}{(2\pi )^3}Tr(W_a\star W_b\star W_c)
	\label{approxinhomoaveragecurrent}
\end{align}

The perturbative non-renormalization theorem proven in \cite{ZZ2021} allows to draw conclusions on the average current in (\ref{approxinhomoaveragecurrent}) in the interacting theories. Take $a\in \{ \eta ,\alpha ,\lambda \}$. Define the renormalized current 

\begin{align}
	{\cal J}^k_{aa}(x)=-\int \frac{d^3p}{(2\pi )^3}Tr(({\cal G}_a)_W(x,p)\frac{\partial}{\partial p_k}({\cal Q}_a)_W(x,p))
\end{align}

with $({\cal Q}_a)_W=(Q_0)_W-(\Sigma_W)_a$, where $\Sigma_a$ is the self-energy

\begin{align}
	\bar{\cal J}^k_{aa}=\frac{1}{\beta L^2}\int d^3x{\cal J}^k_{aa}(x). 
\end{align}

At the same time we have

\begin{align}
	{\cal J}^k_{a}(x)=-\int \frac{d^3p}{(2\pi )^3}Tr(({\cal G}_a)_W(x,p)\frac{\partial}{\partial p_k}({ Q}_0)_W(x,p))
\end{align}

 and 

\begin{align}
	\bar{\cal J}^k_{a}=\frac{1}{\beta L^2}\int d^3x{\cal J}^k_{a}(x). 
\end{align}

In place of (\ref{identityfree}) we further have the identity

\begin{align}
	({\cal G}_a)_W(x,p)\star ({\cal Q}_a)_W(x,p)=1.
	\label{identityinteracting}
\end{align}

in the presence of interactions.

The non-renormalization theorem \cite{ZZ2021} shows that $\bar{\cal J}^k_{aa}=\bar{\cal J}^k_{a}=\bar{J}^k_0$. We now come back to the above Gedankenexperiment. Mark the half incorporating mutual interactions by I and that without mutual interactions by II. The total averaged current ${\cal J}^k_{a}$ of the considered system may be written as 

\begin{align}
	{\cal J}^k_{a}=({\cal J}^I)^k_{persistent,a}+({\cal J}^{II})^k_{persistent,0}+\epsilon_{kl}(\sigma^I_aE_l-\sigma^{II}_0E_l).
\end{align}

By the first two summands we denote persistent current contributions which may already arise in the absence of an external electric field. According to the Bloch theorem the appearance of these terms is not allowed in many relevant systems (those to which it applies). For the Hall conductivity only the last two contributions to the current are of interest, where the proportionality constants are the respective conductivities. The non-renormalization of the current by interactions then implies

\begin{align}
	\sigma^I_{a}=\sigma^{II}_0
\end{align}

and therefore the non-renormalization of the conductivity.

It is important to note that the conductivity derived from ${\cal J}^k_{aa}$ is a topological invariant and not that derived from ${\cal J}^k_{a}$. The proof of the topological invariance underlines this, as we make use of (\ref{identityinteracting}). The proof that $\bar{\cal J}^k_{a}$ is a topological invariant is given in Appendix \ref{AppA}. The similar proof is valid for the expression of $\sigma_a^I$ through the Wigner transformed Green function (see \cite{ZW2019}).  

\begin{align}
	\nonumber \sigma_a^I 	=&\frac{1}{3! (2\pi )^3\beta L^2}\epsilon_{abc}\int d^3 x\int d\omega \int_{BZ}dp_1dp_2 Tr({{\cal G}}^0(\omega ,p_1,p_2)\star\frac{\partial {{\cal Q}}^0(\omega ,p_1,p_2)} {\partial p_a}\star{{\cal G}}^0(\omega ,p_1,p_2)\\&\star \frac{\partial {{\cal Q}}^0(\omega ,p_1,p_2)}{\partial p_b}\star{{\cal G}}^0(\omega ,p_1,p_2)\star \frac{\partial {{\cal Q}}^0(\omega ,p_1,p_2)}{\partial p_c})
\end{align}

\section{Interacting systems. Non - perturbative analysis.}

\subsection{Groenewold equation and its iterative solution for the interacting systems}

Let $\hat {\cal G}$ be the interacting two - point Green function. The renormalized Dirac operator $\hat {\cal Q}$  is defined as a solution of equation
\be \hat {\cal Q} \hat {\cal G}=1 \ee
Weyl-Wigner transformation results in the interacting version of Groenewold equation
\be {\cal Q}_W(x,p) \star  {\cal G}_W(x,p)=1 \label{Groen}\ee

We assume here that all external fields vary slowly, i.e. these variations may be neglected at the distance of the order of lattice spacing. Then Weyl symbol of bare (non - interacting) Dirac operator has the functional dependence  ${ Q}_W(x,p-A(x))$ in the presence of constant external field $A_i(x)$ (corresponding to the field strength $F_{ij}$). Here the coordinate dependence caused by the other external fields is given by direct dependence on $x$.
Function ${\cal Q}_W(x,p)$ with interaction corrections can be represented as
\begin{equation}
	{\cal Q}_W(x,p) = {\cal Q}^{(0)}_W(x,p-A(x)) + {\cal Q}^{(1)}_{(ij) \,W}(x,p-A(x))F_{ij} + ...\label{QpA}
\end{equation}
Dots represent the terms proportional to the higher powers of $F$ and the derivatives of $F$. This expansion is valid under the condition $|\lambda^2 F_{ij}| \ll 1$, where $\lambda$ is the correlation length associated with the given interacting system. This expansion is reasonable, at least, when we consider the DC conductivity, i.e. the response of the  current to sufficiently small external electric field. The correlation length associated with bare Dirac operator is typically equal to the lattice spacing. In the presence of interactions the correlation length may become much larger, of the order of the existing dimensional parameters of the system.

In the similar way the solution of Groenewold equation (\ref{Groen}) for the interacting Green function (up to the terms linear in $F$) is given by
$$
{\cal G}_W(x,p) \approx {\cal G}^{(0)}_W(x,p)+{\cal G}_{(ij)W}^{(1)}F_{ij}
$$
where  ${\cal G}^{(0)}_W(x,p)$ is solution of reduced Groenewold equation (i.e. the one without $A(x)$):
$$
{\cal G}^{(0)}_W(x,p)\star {\cal Q}^{(0)}_W(x,p)=1
$$
The first order term in derivative of $A$ is more complicated than in case of non - interacting Green function:
\begin{eqnarray}
	{\cal G}_{(ij)W}^{(1)}&=&
	\frac{i}{2} \Big[ {\cal G}_W^{(0)}\star \(\pd_{p_i} {\cal Q}_W^{(0)}\) \star {\cal G}_W^{(0)}
	\star \(\pd_{p_j} {\cal Q}_W^{(0)}\) \star {\cal G}_W^{(0)} \Big]F_{ij}\nonumber\\ && - {\cal G}_W^{(0)}\star {\cal Q}_{(ij)W}^{(1)} \star {\cal G}_W^{(0)} F_{ij}
\end{eqnarray}

\subsection{Topological expression for $	{\cal J}^k_{aa}(x)$}

Let us consider the "renormalized" electric current
\begin{align}
	{\cal J}^k_{aa}(x)=-\int \frac{d^3p}{(2\pi )^3}Tr(({\cal G}_a)_W(x,p)\frac{\partial}{\partial p_k}({\cal Q}_a)_W(x,p))
\end{align}
It differs from the precise expression by the substitution of the renormalized velocity operator instead of the bare one. In Appendix \ref{AppB} we prove the non - homogeneous version of Ward identity
\begin{align}
	\Gamma^{\mu}_W(P,x|0)=-i\frac{\partial}{\partial p_{\mu}}{\cal Q}_W(p,x)|_{p=P}.
\end{align} 
Here $\Gamma^{\mu}_W(P,x|k)$ is the (Wigner transformed) renormalized vertex for the emission of photon carrying momentum $k$ by non - homogeneous system of electrons (those that exist in the presence of both external electric and external magnetic fields). The physical meaning of $\Gamma$ is the renormalized velocity operator. Therefore, we can represent ${\cal J}^k_{aa}(x)$ as
 \begin{align}
 	{\cal J}^k_{aa}(x)=\int \frac{d p^3_1}{(2\pi)^3}\frac{d p^3_2}{(2\pi)^3}e^{i (p_1 - p_2)x}\langle \bar{\Psi}(p_1) \Gamma^k(p_1,p_2,0) \Psi(p_2) \rangle
 \end{align}

We obtain the following term with the linear response to external field strength:
\be
{\cal J}^k_{aa}(x)&=-
\frac{i}{2}
\int_{\mathcal M} \frac{d^3p}{(2\pi)^3}
\tr \Bigl[
\Big[ {\cal G}_W^{(0)}\star \(\pd_{p_i} {\cal Q}_W^{(0)}\) \\&\star {\cal G}_W^{(0)}
\star \(\pd_{p_j} {\cal Q}_W^{(0)}\) \star {\cal G}_W^{(0)} \Big] \pd_{p_k}{\cal Q}_W^{(0)}
\Bigr]
F_{ij}\\
& + \int_{\mathcal M} \frac{d^Dp}{(2\pi)^D}
\tr \Bigl[
\Big[ {\cal G}_W^{(0)}\star {\cal Q}_{(ij)W}^{(1)} \star {\cal G}_W^{(0)}  \Big] \pd_{p_k}{\cal Q}_W^{(0)}
\Bigr]
F_{ij}
\\
& - \int_{\mathcal M} \frac{d^Dp}{(2\pi)^D}
\tr \Bigl[ {\cal G}_W^{(0)}   \pd_{p_k} \Big[{\cal Q}_{(ij)W}^{(1)}\Big]
\Bigr]
F_{ij}
\label{ji5lr}\ee
Averaging the local current over the whole system volume we get
\be
\bar{\cal J}^k_{aa}&=
-\frac{i}{2}\frac{1}{\beta{\bf V}}
\int d^3x\int_{\mathcal M} \frac{d^3p}{(2\pi)^3}\tr \Bigl[
\Bigl[ {\cal G}_W^{(0)}\star \(\pd_{p_i} {\cal Q}_W^{(0)}\) \\&\star {\cal G}_W^{(0)}
\star \(\pd_{p_j} {\cal Q}_W^{(0)}\) \star {\cal G}_W^{(0)}\pd_{p_k}{\cal Q}_W^{(0)} \\&
+ 2i {\cal G}_W^{(0)}\star {\cal Q}_{(ij)W}^{(1)} \star {\cal G}_W^{(0)} \pd_{p_k}{\cal Q}_W^{(0)}
\\& - 2i  {\cal G}_W^{(0)}   \pd_{p_k} \Big[{\cal Q}_{(ij)W}^{(1)}\Big] \Bigr]
\Bigr]
F_{ij}
\label{Ji5lr}\ee
In the above expression the star product may be inserted. As a result the last two terms cancel each other, and we are left with
$$
\bar{\cal J}^k_{aa} = \epsilon^{kl} \tilde{\sigma}_H E_l
$$
where $E_l$ is electric field while
\be
\tilde{\sigma}_H&=
\frac{1}{6}\frac{\epsilon^{ijk}}{\beta{\bf V}}
\int d^3x\int_{\mathcal M} \frac{d^3p}{(2\pi)^3}\tr  {\cal G}_W^{(0)}\star \(\pd_{p_i} {\cal Q}_W^{(0)}\) \\&\star {\cal G}_W^{(0)}
\star \(\pd_{p_j} {\cal Q}_W^{(0)}\) \star {\cal G}_W^{(0)}\star \pd_{p_k}{\cal Q}_W^{(0)} \label{sigmaH5}\ee

\subsection{Difference between  $	{\cal J}^k_{aa}(x)$ and $	{\cal J}^k_{a}(x)$ }

The original expression for electric current (with the non - renormalized velocity operator) results in 
$$
\bar{\cal J}^k_{a} = \epsilon^{kl} {\sigma}_H E_l
$$
where 
\be
{\sigma}_H&=
\frac{1}{6}\frac{\epsilon^{ijk}}{\beta{\bf V}}
\int d^3x\int_{\mathcal M} \frac{d^3p}{(2\pi)^3}\tr  {\cal G}_W^{(0)}\star \(\pd_{p_i} {\cal Q}_W^{(0)}\) \\&\star {\cal G}_W^{(0)}
\star \(\pd_{p_j} {\cal Q}_W^{(0)}\) \star {\cal G}_W^{(0)}\star \pd_{p_k}{Q}_W^{(0)} \ee

The difference with the expression for $\tilde{\sigma}_H$ is in the bare Dirac operator $Q$ standing instead of $\cal Q$.

The difference between the two expressions is
\be
{\sigma}_H - \tilde{\sigma}_H&=
\frac{1}{6}\frac{\epsilon^{ijk}}{\beta{\bf V}}
\int d^3x\int_{\mathcal M} \frac{d^3p}{(2\pi)^3}\tr  {\cal G}_W^{(0)}\star \(\pd_{p_i} {\cal Q}_W^{(0)}\) \\&\star {\cal G}_W^{(0)}
\star \(\pd_{p_j} {\cal Q}_W^{(0)}\) \star {\cal G}_W^{(0)}\star \pd_{p_k}{\Sigma}_W^{(0)} \ee 

For the electron self energy we have the Schwinger - Dyson equation:
\begin{eqnarray}
	&&	\Sigma_W(x,z) = \int d^3x^\prime d^3z^\prime {\cal D}(x,z^\prime) \,  {\cal G}^{(0)}(x,x^\prime)  \Gamma(x^\prime,z|z^\prime) 
\end{eqnarray}
Here ${\cal D}^{}_{}(x,z^\prime)$ is the complete interacting propagator of an excitation that provides interactions.   

In practise a certain non - perturbative treatment of the problem may be achieved already in the rainbow approximation, when the renormalized vertex is replaced by the bare one while the renormalized propagator of scalar excitation is replaced by the bare propagator. 
\begin{eqnarray}
	&&	\Sigma_W(x,p) \approx \int \frac{d^3k}{(2\pi)^3} { D}^{}_{W}(k) \,  {\cal G}^{(0)}_W(x,p-k) 
\end{eqnarray}
In this approximation we obtain (in the presence of both magnetic and electric fields)
\begin{eqnarray}
	{\cal J}^k_{a}(x) - {\cal J}^k_{aa}(x)&\approx&-
		\int_{\mathcal M} \frac{d^3p}{(2\pi)^3} \tr  {\cal G}_W^{}  \pd_{p_k}{\Sigma}_W^{}\nonumber\\
		&=&
		\int_{\mathcal M} \frac{d^3p}{(2\pi)^3} \tr  \pd_{p_k}{\cal G}_W^{}  {\Sigma}_W^{}\nonumber\\
		 &=&
	\int_{\mathcal M} \frac{d^3p}{(2\pi)^3}\frac{d^3k}{(2\pi)^3}\tr  \pd_{p_k}{\cal G}_W^{}(x,p) {\cal G}^{}_W(x,p-k) {D}^{}_{W}(k) 
	\nonumber\\
	&=&
	\int_{\mathcal M} \frac{d^3\bar{p}}{(2\pi)^3}\frac{d^3k}{(2\pi)^3}\tr  \pd_{\bar{p}_k}{\cal G}_W^{}(x,\bar{p}+k){D}^{}_{W}(k) {\cal G}^{}_W(x,\bar{p}) 
		\nonumber\\
	&=&
	\int_{\mathcal M} \frac{d^3\bar{p}}{(2\pi)^3}\frac{d^3\bar{k}}{(2\pi)^3}\tr  \pd_{\bar{p}_k}{\cal G}_W^{}(x,\bar{p}-\bar{k}){D}^{}_{W}(-\bar{k}) {\cal G}^{}_W(x,\bar{p}) 
	\nonumber\\
	&=&
	\int_{\mathcal M} \frac{d^3\bar{p}}{(2\pi)^3}\frac{d^3\bar{k}}{(2\pi)^3}\tr  \pd_{\bar{p}_k}{\cal G}_W^{}(x,\bar{p}-\bar{k}){D}^{}_{W}(\bar{k}) {\cal G}^{}_W(x,\bar{p}) \nonumber\\ &=&
	\int_{\mathcal M} \frac{d^3p}{(2\pi)^3} \tr  {\cal G}_W^{}  \pd_{p_k}{\Sigma}_W^{}   
\end{eqnarray}
This expression vanishes identically, and as a result the two current densities $	{\cal J}^k_{a}(x)$ and $ {\cal J}^k_{aa}(x)$ coincide in this approximation. As a result also the responses of those currents to external electric field coincide, and we come to the coincidence of $\sigma_H$ and $\tilde{\sigma}_H$ in this approximation.

Now let us come to the approximation, when renormalized vertex $\Gamma$ is replaced by the bare one, while scalar boson propagator remains complete:
\begin{eqnarray}
	&&	\Sigma_W(x,p) \approx \int \frac{d^3k}{(2\pi)^3} {\cal D}^{}_{W}(x,k) \,  {\cal G}^{(0)}_W(x,p-k) 
\end{eqnarray}
This approximation is much more powerfull than the rainbow approximation, and we suppose that it is able to catch the basic nonperturbative properties of Coulomb interactions, including those that lead to formation of the FQHE. 
In this approximation 
\begin{eqnarray}
	{\cal J}^k_{a}(x) - {\cal J}^k_{aa}(x)&\approx&
	-\int_{\mathcal M} \frac{d^3p}{(2\pi)^3} \tr  {\cal G}_W^{}  \pd_{p_k}{\Sigma}_W^{}\nonumber\\
	&=&
		\int_{\mathcal M} \frac{d^3p}{(2\pi)^3} \tr  \pd_{p_k}{\cal G}_W^{}  {\Sigma}_W^{}\nonumber\\
	&=&
	\int_{\mathcal M} \frac{d^3p}{(2\pi)^3}\frac{d^3k}{(2\pi)^3}\tr  \pd_{p_k}{\cal G}_W^{}(x,p) {\cal G}^{}_W(x,p-k) {\cal D}^{}_{W}(x,k) 
	\nonumber\\
	&=&
	\int_{\mathcal M} \frac{d^3\bar{p}}{(2\pi)^3}\frac{d^3k}{(2\pi)^3}\tr  \pd_{\bar{p}_k}{\cal G}_W^{}(x,\bar{p}+k){\cal D}^{}_{W}(x,k) {\cal G}^{}_W(x,\bar{p}) 
	\nonumber\\
	&=&
	\int_{\mathcal M} \frac{d^3\bar{p}}{(2\pi)^3}\frac{d^3\bar{k}}{(2\pi)^3}\tr  \pd_{\bar{p}_k}{\cal G}_W^{}(x,\bar{p}-\bar{k}){\cal D}^{}_{W}(x,-\bar{k}) {\cal G}^{}_W(x,\bar{p}) 
\end{eqnarray}
The difference with the case of rainbow approximation is the presence of dependence of ${\cal D}^{}_{W}(x,{k}) $ on $x$. 

Let $\phi$ represent the real - valued field, which is responsible for the interactions. Coulomb interactions, contact four - fermion interactions and Yukawa interactions, as well as a lot of other types of interactions are described by such field. Then
\begin{eqnarray}
{\cal D}^{}_{W}(x,{k}) = \frac{1}{Z}\int d^3y e^{-iyk}\int D\phi D\bar{\psi} D\psi e^{-S[\bar{\psi},\psi,\phi]}\phi(x+y/2)\phi(x-y/2)  	
\end{eqnarray}
Here $S$ is the action depending on the fields. One can see that 
 \begin{eqnarray}
 	{\cal D}^{}_{W}(x,-{k}) &=& \frac{1}{Z}\int d^3y e^{iyk}\int D\phi D\bar{\psi} D\psi e^{-S[\bar{\psi},\psi,\phi]}\phi(x+y/2)\phi(x-y/2) \nonumber\\ & = & 
 	 \frac{1}{Z}\int d^3y e^{-iyk}\int D\phi D\bar{\psi} D\psi e^{-S[\bar{\psi},\psi,\phi]}\phi(x-y/2)\phi(x+y/2) =  {\cal D}^{}_{W}(x,{k})	
 \end{eqnarray}
Therefore, we obtain
\begin{eqnarray}
	{\cal J}^k_{a}(x) - {\cal J}^k_{aa}(x)&\approx&
	-\int_{\mathcal M} \frac{d^3p}{(2\pi)^3} \tr  {\cal G}_W^{}  \pd_{p_k}{\Sigma}_W^{}\nonumber\\
	 &=&
	\int_{\mathcal M} \frac{d^3p}{(2\pi)^3} \tr  {\cal G}_W^{}  \pd_{p_k}{\Sigma}_W^{} = 0  
\end{eqnarray}
We come to conclusion that in this approximation as well  the value of $\sigma_H$ does not differ from $\tilde{\sigma}_H$. Notice that although this is still an approximation, it is already out of the perturbation theory. The value of Hall conductivity is then given by Eq.  (\ref{sigmaH5}).

\section{Transition to Harper representation in interacting systems}

Next, we will rewrite the momentum space integration in Eq. (\ref{sigmaH5}) by introducing magnetic Brillouin zones (as in Sect. \ref{SectHarper}). The goal will be to absorb the magnetic field within the Harper representation. This is due to the fact that the constant magnetic field in a finite sized system is quantized according to $B=2\pi\frac{\nu}{N}$ with mutually simple integers $\nu$ and $N$. The introduction of magnetic Brillouin zone using an example of $1D$ system is given in details in Appendix \ref{AppD}. It may be extended in a straightforward way to the 2D systems.

The interacting theories were considered for inhomogeneous (magnetic) field configurations within the Wigner-Weyl formalism in \cite{ZW2019}. The treatment of homogeneous systems, which have been discussed in Harper representation in Sect. \ref{SectHarper}, are therefore just a special case of this setup. We may therefore apply the non-renormalization theorem also for the current and conductivities expressed in Harper representation. This is done in the following. We remark that each Moyal star product may be replaced by ordinary multiplication in the expressions derived within the inhomogeneous framework. The Wigner transformed propagator as well as its inverse lack explicit spatial dependence in the case of a spatially homogeneous system. This is because their momentum space matrix elements become diagonal. We will finally write both the averaged current and the conductivity in Harper representation. Once the transition to Harper representation is made, the magnetic field will have been absorbed completely into the matrix structure. We will mark those matrices by using bold letters. The matrices in Harper representation then do not inherit diagonality as do the continuous variables because of the matrix index shift induced by the magnetic field. Since this rewriting is identical for both free and interacting theories we again omit the subscript used to distinguish between them. We will write the expression at zero temperature only.

Below we omit index $a$ for brevity. The averaged current may be rewritten as follows

\begin{align}
	\nonumber \bar{\cal J}^k=&-\frac{1}{TL^2}\int d^3x\frac{d^3p}{(2\pi )^3}Tr({\cal G}_W(x,p)\star \frac{\partial {\cal Q}_W(x,p)}{\partial p_k})\\
	\nonumber =&+\frac{1}{TL^2}\int d^3x\frac{d^3p}{(2\pi )^3}Tr({\cal G}_W(x,p)\star \frac{\delta {\cal Q}_W(x,p)}{\delta A_k})\\
	\nonumber =&+\frac{1}{TL^2}\int d^3x\frac{d^3p}{(2\pi )^3}Tr((\hat{{\cal G}}\frac{\delta \hat{{\cal Q}}}{\delta A_k})_W)\\
	\nonumber =&+\frac{1}{TL^2}Tr(\hat{{\cal G}}\frac{\delta \hat{{\cal Q}}}{\delta A_k})\\
	\nonumber =&+\frac{1}{TL^2}\int d^3p^{\prime}d^3p Tr({\cal G}(p^{\prime},p)\frac{\delta {\cal Q}(p,p^{\prime})}{\delta A_k})\\
	\nonumber =&-\frac{1}{TL^2}\int d^3p^{\prime}d^3pTr({\cal G}(p^{\prime},p)\Big(\frac{\partial }{\partial p_k}+\frac{\partial }{\partial p^\prime_k}\Big){\cal Q}(p,p^{\prime}))\\
	\nonumber =&-\frac{1}{TL^2}\int d\omega^{\prime}d\omega \int_{BZ} dp^{\prime}_1dp_1\int_{BZ}dp^{\prime}_2dp_2Tr({\cal G}(\omega^{\prime},p^{\prime}_1,p^{\prime}_2,\omega ,p_1,p_2)\Big(\frac{\partial }{\partial p_k}+\frac{\partial }{\partial p^\prime_k}\Big){\cal Q}(\omega ,p_1,p_2,\omega^{\prime},p^{\prime}_1,p^{\prime}_2)
	\nonumber\\ =&-\frac{1}{TL^2}\int d\omega d\omega^\prime \sum_{n^{\prime},n=0}^{N-1}\int_{\frac{BZ}{N}} dk^{\prime}_1dk_1\int_{BZ}dk^{\prime}_2dp_2Tr({\cal G}_{n^{\prime}n}(\omega^{\prime},k^{\prime}_1,k^{\prime}_2,\omega ,k_1,k_2)\Big(\frac{\partial }{\partial k_k}+\frac{\partial }{\partial k^\prime_k}\Big){\cal Q}_{n n^\prime}(\omega ,k_1,k_2,\omega^{\prime},k^{\prime}_1,k^{\prime}_2))\\
	\nonumber =&-\frac{1}{(2\pi )^3}\int d\omega\int_{\frac{BZ}{N}} dk_1\int_{BZ}dp_2Tr(\textbf{{\cal G}}(\omega ,k_1,p_2)\frac{\partial \textbf{{\cal Q}}(\omega ,k_1,p_2)}{\partial p_k})\\
	=&-\frac{1}{(2\pi )^3N}\int d\omega\int_{BZ} dp_1dp_2Tr(\textbf{{\cal G}}(\omega ,p_1,p_2)\frac{\partial \textbf{{\cal Q}}(\omega ,p_1,p_2)}{\partial p_k}).\label{QQQ}
\end{align}

Here the results of Appendix \ref{AppD} are used in order to replace an integral over the Brillouin zone by the integral over the magnetic Brillouin zone and the sum over the integer number that runs from $1$ to $N$. Finally, the integral over each  magnetic Brilloin zone is extended to the whole Brillouin zone, while the result is divided by $N$. We also have 

$$
{\cal Q}_{n n^\prime}(\omega ,k_1,k_2,\omega^{\prime},k^{\prime}_1,k^{\prime}_2)) = \delta^{(2)}(k - k^\prime)\delta(\omega - \omega^\prime)\textbf{{\cal Q}}_{n n^\prime}(\omega ,k_1,k_2)
$$
and
$$
\Big(\frac{\partial }{\partial k_k}+\frac{\partial }{\partial k^\prime_k}\Big) {\cal Q}_{n n^\prime}(\omega ,k_1,k_2,\omega^{\prime},k^{\prime}_1,k^{\prime}_2)) = \delta^{(2)}(k - k^\prime)\delta(\omega - \omega^\prime)\frac{\partial }{\partial k_k}\textbf{{\cal Q}}_{n n^\prime}(\omega ,k_1,k_2)
$$

Note that to get to the last equality of Eq. (\ref{QQQ}) from the previous line we used that integration over different magnetic Brillouin zones induces a shift in the Harper matrices which is rendered equal to the unshifted case due to the trace. Thus integration is equal for each magnetic Brillouin zone. The diagonality of both propagator and its inverse in momentum space induce two delta functions. Half of them are being used to reduce the number of integrations by half. This leaves behind coincident delta functions which are regularized by the finite system size according to 

\begin{align}
	\delta (p-p)=\frac{TL^2}{(2\pi )^3}.
\end{align}

We may employ the same manipulations for the conductivity as those used for the current. The conductivity may be brought into Harper representation by the following sequence of equalities

\begin{align}
	\nonumber \sigma_H=&\frac{1}{3!}\epsilon_{abc}\int\frac{d^2x}{L^2}\frac{d^3p}{(2\pi )^3}Tr(W_a\star W_b\star W_c)\\
	\nonumber =&\frac{1}{3!}\epsilon_{abc}\int\frac{d^3x}{TL^2}\frac{d^3p}{(2\pi )^3}Tr({\cal G}^0_W(x,p)\star \partial_{p_a}{\cal Q}^0_W(x,p)\star {\cal G}^0_W(x,p)\star \partial_{p_b}{\cal Q}^0_W(x,p)\star {\cal G}^0_W(x,p)\star \partial_{p_c}{\cal Q}^0_W(x,p))\\
	\nonumber =&-\frac{1}{3!}\epsilon_{abc}\int\frac{d^3x}{TL^2}\frac{d^3p}{(2\pi )^3}Tr({\cal G}^0_W(x,p)\star \frac{\delta}{\delta A_a}{\cal Q}^0_W(x,p)\star {\cal G}^0_W(x,p)\star \frac{\delta}{\delta A_b}{\cal Q}^0_W(x,p)\star {\cal G}^0_W(x,p)\star \frac{\delta}{\delta A_c}{\cal Q}^0_W(x,p))\\
	\nonumber =&-\frac{1}{3!}\epsilon_{abc}\int \frac{d^3x}{TL^2}\frac{d^3p}{(2\pi )^3}Tr((\hat{{\cal G}}^0\frac{\delta \hat{{\cal Q}}^0}{\delta A_a}\hat{{\cal G}}^0\frac{\delta \hat{{\cal Q}}^0}{\delta A_b}\hat{{\cal G}}^0\frac{\delta \hat{{\cal Q}}^0}{\delta A_c})_W)\\
	\nonumber =&-\frac{1}{3!TL^2}\epsilon_{abc}Tr(\hat{{\cal G}}^0\frac{\delta \hat{{\cal Q}}^0}{\delta A_a}\hat{{\cal G}}^0\frac{\delta \hat{{\cal Q}}^0}{\delta A_b}\hat{{\cal G}}^0\frac{\delta \hat{{\cal Q}}^0}{\delta A_c})\\
	\nonumber =&-\frac{1}{3!TL^2}\epsilon_{abc}\int d^{18}p_{uvwxyz}Tr({{\cal G}}^0(p_u,p_v)\frac{\delta {\cal Q}^0(p_v,p_w)}{\delta A_a}{\cal G}^0(p_w,p_x)\frac{\delta {\cal Q}^0(p_x,p_y)}{\delta A_b}{\cal G}^0(p_y,p_z)\frac{\delta {\cal Q}^0(p_z,p_u)}{\delta A_c})\\
	\nonumber =&-\frac{1}{3!TL^2}\epsilon_{abc}\int^{\prime} d^{18}p_{uvwxyz}Tr(\textbf{{\cal G}}^0(p_u,p_v)\frac{\delta \textbf{{\cal Q}}^0(p_v,p_w)}{\delta A_a}\textbf{{\cal G}}^0(p_w,p_x)\frac{\delta \textbf{{\cal Q}}^0(p_x,p_y)}{\delta A_b}\textbf{{\cal G}}^0(p_y,p_z)\frac{\delta \textbf{{\cal Q}}^0(p_z,p_u)}{\delta A_c})\\
	\nonumber =&-\frac{1}{3!(2\pi )^3}\epsilon_{abc}\int d\omega \int_{\frac{BZ}{N}}dk_1\int_{BZ}dp_2 Tr(\textbf{{\cal G}}^0(\omega ,k_1,p_2)\frac{\delta \textbf{{\cal Q}}^0(\omega ,k_1,p_2)}{\delta A_a}\textbf{{\cal G}}^0(\omega ,k_1,p_2)\frac{\delta \textbf{{\cal Q}}^0(\omega ,k_1,p_2)}{\delta A_b}\textbf{{\cal G}}^0(\omega ,k_1,p_2)\frac{\delta \textbf{{\cal Q}}^0(\omega ,k_1,p_2)}{\delta A_c})\\
	\nonumber =&\frac{1}{3!(2\pi )^3}\epsilon_{abc}\int d\omega \int_{\frac{BZ}{N}}dk_1\int_{BZ}dp_2 Tr(\textbf{{\cal G}}^0(\omega ,k_1,p_2)\frac{\partial \textbf{{\cal Q}}^0(\omega ,k_1,p_2)}{\partial p_a}\textbf{{\cal G}}^0(\omega ,k_1,p_2)\frac{\partial \textbf{{\cal Q}}^0(\omega ,k_1,p_2)}{\partial p_b}\textbf{{\cal G}}^0(\omega ,k_1,p_2)\frac{\partial \textbf{{\cal Q}}^0(\omega ,k_1,p_2)}{\partial p_c})\\
	=&\frac{1}{3!N(2\pi )^3}\epsilon_{abc}\int d\omega \int_{BZ}dp_1dp_2 Tr(\textbf{{\cal G}}^0(\omega ,p_1,p_2)\frac{\partial \textbf{{\cal Q}}^0(\omega ,p_1,p_2)}{\partial p_a}\textbf{{\cal G}}^0(\omega ,p_1,p_2)\frac{\partial \textbf{{\cal Q}}^0(\omega ,p_1,p_2)}{\partial p_b}\textbf{{\cal G}}^0(\omega ,p_1,p_2)\frac{\partial \textbf{{\cal Q}}^0(\omega ,p_1,p_2)}{\partial p_c})\label{sigmaH6}
\end{align}

The superscripts $0$ signify $E=0$ in both propagator and inverse propagator. The prime at the integral marks a different integration region, namely that over the magnetic Brillouin zone for the first spatial momenta. The diagonality of both propagator and inverse propagator cancel five of the six momentum integrations, while the sixth delta function cancels the volume factor in the denominator in the same way as for the current. Note that again the cancellation only happens for the continuous integrations, as the magnetic field induces non-diagonality in the Harper matrices.

The non-renormalization of the first expressions in the two sequences of equalities then trivially imply that of the last one in Harper representation.

\section{Conclusions and discussion}
\label{SectionV}

In the present paper we considered the fermionic systems in two space dimensions. First we restrict our consideration to the non - interacting systems defined on rectangular lattices. The consideration is illustrated by the model with the simplest Hamiltonian of Eq. (\ref{Hsimple}). However, our results are valid for the tight - binding Hamiltonian of a more general type.

We consider the given systems in the presence of constant external magnetic field such that the magnetic flux through the lattice cell is equal to
$$
\Phi = \frac{\nu}{N} \times \Phi_0
$$
where $\Phi_0$ is a quantum of magnetic flux $\Phi_0 = h/e$.

We divide the Brillouin zone into the Magnetic Brillouin zones with eigenvectors of momentum given by Eq. (\ref{HarperV}). Harper representation of Hamiltonian is its representation in the basis of Eq. (\ref{HarperV}). In this representation Hamiltonian $\hat{H}$ is diagonal in $u$ and $v$ but is not diagonal in $n$. For the simplest tight - binding model of Eq. (\ref{Hsimple}) the lattice Dirac operator ${\bf Q} = -i\om + \hat{H}$ in Harper representation has nonzero matrix elements  with the transitions between the adjacent values of $n$. In a more general case the lattice Dirac operator in Harper representation has the form with the transitions between any possible pairs of $n$, but remains diagonal in $u$ and $v$. Dirac operator $\bf Q$ becomes the $N\times N$ matrix depending on $u$ and $v$. The Green function ${\bf G}= {\bf Q}^{-1}$ also becomes the $N\times N$ matrix. Both $\bf Q$ and $\bf G$ defined in this way belong to the magnetic Brillouin zone,  which is $N$ times smaller than the whole Brillouin zone, i.e. $u \in [\Delta, \Delta + \frac{2 \pi }{a N})$, while $v \in [0, \frac{2 \pi }{a })$. It is worth mentioning that matrix $\bf Q$ obeys specific boundary conditions that depend on $n$.

However, we are able to extend the definition of matrices $\bf Q$ and $\bf G$ to the whole Brillouin zone. After this extension matrix $\bf Q$ obeys periodic boundary conditions in the whole Brillouin zone. As a result we are able to represent the Hall conductivity of the given system as 
\begin{equation}
\boxed{\sigma_H = \frac{e^2}{h} \, \frac{1}{N}\,{\cal N}}\label{sigmaH7}
 \end{equation} 
Here $\cal N$ is the topological invariant composed of the $N\times N$ matrices $\bf Q$ and $\bf G$:
\begin{equation}
\boxed{	{\cal N} 	=\frac{1}{3!(2\pi )^2}\epsilon_{abc}\int d\omega \int_{BZ}dp_1dp_2 Tr(\textbf{{\cal G}}(\omega ,p_1,p_2)\frac{\partial \textbf{{\cal Q}}(\omega ,p_1,p_2)}{\partial p_a}\textbf{{\cal G}}(\omega ,p_1,p_2)\frac{\partial \textbf{{\cal Q}}(\omega ,p_1,p_2)}{\partial p_b}\textbf{{\cal G}}(\omega ,p_1,p_2)\frac{\partial \textbf{{\cal Q}}(\omega ,p_1,p_2)}{\partial p_c})}\label{N100}
\end{equation} 

Notice that this expression is valid for the tight - binding fermionic system of general form. For the non - interacting systems this invariant {\it should} be equal to the integer multiple of $N$ in order to provide the integer QHE. We check this numerically for the cases with $\nu = 1$, $N = 3,4$. The values of Hall conductivity obtained using solution of Diophantine equation are reproduced as it should be.

The experience of the previously known expressions for the Hall conductivity indicates that Eq. (\ref{HC}) remains valid also for the case of an interacting system. Then matrix $\bf G$ is likely to be replaced by the complete electron propagator in Harper representation (defined originally in the magnetic Brillouin zone, but extended analytically to the whole Brillouin zone) while ${\bf Q} = {\bf G}^{-1}$. This is justified partially by the results obtained by the present authors (see \cite{ZZ2019_0,ZZ2019_1,ZZ2021}). Namely, we have shown that the topological invariant similar to the one of our present Eq. (\ref{HC}) still represents the integer QHE conductivity in the presence of interactions if the Green function is substituted by the interacting one. The very method used in \cite{ZZ2019_0,ZZ2019_1,ZZ2021} may be extended to prove the similar statement for our Eq. (\ref{HC}).

Thus we extend our consideration to the interacting systems. Here we use the so - called approximate version of Wigner - Weyl calculus \cite{ZZ2021}, which works perfectly for the realistic solid state systems with any form of the crystal lattice in the presence of realistic external magnetic fields much smaller than $10^5$ Tesla. First of all we recall the results of \cite{ZZ2021}, where it has been proven that expression of Eq. (\ref{sigmaH5}) for the Hall conductivity remains valid when we substitute the complete interacting Green function instead of the non - interacting one. This result has been obtained using perturbation theory, and is valid to all orders. However, also in \cite{ZZ2021} it has been proven that if the interactions are taken into account perturbatively, this value remains equal to the one of the non - interacting theory. Therefore, we conclude that the FQHE cannot be described using  perturbation theory - its origin is non - perturbative.

We then consider the interactions non - perturbatively based on the Schwinger - Dyson equations. First we prove that Eq. (\ref{sigmaH5}) remains valid in the so - called rainbow approximation. Next, we prove this statement in more involved approximation, when both fermion and scalar excitation propagators in the Schwinger - Dyson equation are complete non - interacting ones, while the complete one - particle irreducible three point Green function $\Gamma^{}$ is replaced by bare interaction vertex.  Thus we conclude that even the non - perturbative treatment leads us to the obtained representation for Hall conductivity. Next, we prove that in the case of the clean system in the presence of constant magnetic field it is reduced to the one of Eq. (\ref{sigmaH6}), i.e. to Eq. (\ref{sigmaH7}) with $\cal N$ given by  Eq. (\ref{N100}), which is the topological invariant composed of the {\it interacting} Green function taken in Harper representation. Now its value is not necessarily proportional to $N$, and we may, in principle, arrive at the {\it topological description of Fractional Quantum Hall effect} because the obtained value of conductivity is given by the rational number times inverse Klitzing constant. The denominator $N$ of the given rational number may, in principle, take any integer value depending on the value of  magnetic flux through the lattice cell. In practise in the sufficiently clean system at constant value of external magnetic field the value of Hall conductivity should depend on the value of chemical potential. Without interactions this dependence has the step - like form with plateaus representing the integer QHE conductivity (in the case when $\mu$ is kept constant while magnetic field varies the plateaus are not seen unless the disorder is added \cite{Tong:2016kpv}). In the presence of interactions at constant magnetic field the dependence of $\sigma_H$ on $\mu$ at constant $B$ will acquire plateaus corresponding to fractional QHE. (If $\mu$ is fixed while $B$ varies, the fractional plateaus appear in the presence of disorder. One observes more fractional plateaus when the disorder is reduced.) These are the values of $\sigma_H$ proportional to the ratio ${\cal N}/N$. The values of $\cal N$ should depend on $\mu$ and $B$ stepwise, while the values of $N$ are fixed by the choice of $B$ according to Eq. (\ref{NPhi}). Each step in $\cal N$ corresponds to the topological phase transition caused by nonperturbative effects of interactions. The microscopic origin of these transitions remains a miracle to be investigated separately. It would be interesting to calculate the values of $\cal N$ directly within the existing phenomenological models of FQHE (see, for example, \cite{Fradkin,Tong:2016kpv,PhysRevLett.50.1395,PhysRevLett.63.199,PhysRevB.41.7653,Dirac1931}.
This is, however, out of the scope of the present paper.  

We would like to notice the previously proposed expressions for the integer QHE conductivity in the non - interacting systems in terms of the Green functions. In \cite{1r,2r,3r} the interacting models of topological insulators have been considered. Actually, the expressions for the QHE conductivity  of \cite{1r,2r,3r} are similar to our expression, but they are not written in Harper representation and, therefore, are more simple.  Expressions of \cite{4r,5r} differ essentially from our formula, but they  also refer to the models of topological insulators without external magnetic field. As it was mentioned above, our expression works for the system in the presence of external magnetic field. The corresponding systems are non - homogeneous, and, therefore, their consideration is more complicated.   

Actually, our expression of Eq. (\ref{HC}) resembles the one proposed long time ago in \cite{Imai:1990zz} for the systems in the presence of uniform magnetic field. However, the important advantage of our expression is that ${\bm G}_{p_1p_2}^\om$ is $N\times N$ matrix while the Green function entering topological expression of \cite{Imai:1990zz} is infinite dimensional.

M.A.Z. is grateful to G.Volovik for useful discussions, which became the starting point of the present work. M.S. is grateful to I.Fialkovsky for numerous useful private communications.

\appendix

\section{Introduction of the magnetic Brillouin zone}

\label{AppD}

For a single particle 1D lattice we have in momentum space

\begin{align}
	\label{magbri}
	1&=\int_{\Delta}^{\Delta+2\pi}dk|k\rangle\langle k|=\sum_{n=0}^{N-1}\int_{\Delta +n\frac{2\pi}{N}}^{\Delta +(n+1)\frac{2\pi}{N}}dk |k\rangle \langle k|=\sum_{n=0}^{N-1}\int_{\Delta}^{\Delta +\frac{2\pi}{N}}dk |k+n\frac{2\pi}{N}\rangle \langle k+n\frac{2\pi}{N}|\\
	&\equiv \sum_{n=0}^{N-1}\int_{\Delta}^{\Delta +\frac{2\pi}{N}}dk |k,n \rangle \langle k,n|=\sum_{n=0}^{N-1}\int_{\frac{BZ}{N}}dk |k,n \rangle \langle k,n|
\end{align} 

where the first Brillouin zone may be defined as

\begin{align}
	k\in [\Delta ,\Delta +2\pi ]=BZ
\end{align}

with arbitrary $\Delta$. Discrete momentum corresponding to any value of $n$ may be represented as 

\begin{align}
	\frac{2\pi}{N}n=\frac{2\pi}{N}\nu m-2\pi M \Leftrightarrow n=\nu m-NM
	\label{algebra}
\end{align}

where $\nu$ and $N$ are mutually simple positive numbers and $m,n\in \{ 0,...,N-1\}$. We will now show that the assignment $n\mapsto m(n)$ is a bijection.\\ We may assume without loss of generality that $1\leq \nu\leq N-1$, since the unique representation of any $\nu$ by $\nu =aN+b$ with $a\in\mathbb{N}$ and $b\in\{0,...,N-1\}$ and nonzero $a$ may be reduced by sending $M\to M+a$ to the case $a=0$. Define first a map $m\mapsto M(m)$ by the requirement $0\leq\nu m-M(m)N\leq N-1$. Then define a map $m\mapsto n(m)$ by $n(m)=\nu m-M(m)N$. This latter map is injective. If not we could find $m_1\neq m_2$ such that

\begin{align}
	n=\nu m_1-M(m_1)N=\nu m_2-M(m_2)N \Leftrightarrow \nu (m_1-m_2)=N(M(m_1)-M(m_2)).
\end{align}

Without loss of generality (after division) we may assume that $|m_1-m_2|$ and $|M(m_1)-M(m_2)|$ are mutually simple numbers. All prime factors of $N$ are therefore required to be prime factors of $|m_1-m_2|$ as well such that $|m_1-m_2|\geq N$. But we know that by the requirement $m_1,m_2\in\{0,...,N-1\}$ it follows that $|m_1-m_2|<N$, which is a contradiction. This proves injectivity. Surjectivity follows since image and preimage space share the same finite size. This implies bijectivity and as well bijectivity of the inverse. 

Periodicity in momentum space allows to read the very right hand side of (\ref{algebra}) modulo $N$ and therefore we may perform the replacement $n\to\nu m$ as well as to sum over $m$ (which we will rename $n$) in (\ref{magbri}). We then obtain the decomposition of the identity in the form

\begin{align}
	1=\sum_{n=0}^{N-1}\int_{\Delta}^{\Delta +\frac{2\pi}{N}}dk |k+n\frac{2\pi}{N}\nu \rangle \langle k+n\frac{2\pi}{N}\nu |.
\end{align}

In two dimensions and under the assumption of being in Landau gauge we introduce the notation $f(p_1,p_2)\equiv f_n(k_1,p_2)$ as well as $F(p^{\prime}_1,p^{\prime}_2,p_1,p_2)\equiv F_{n^{\prime}n}(k^{\prime}_1,p^{\prime}_2,k_1,p_2)$ for arbitrary functions or matrix elements in momentum space with $p_i,p^{\prime}_i\in BZ$, $k^{\prime}_1=p^{\prime}_1 \, mod \, \frac{BZ}{N}$, $k_1=p_1 \, mod \, \frac{BZ}{N}$ and $p^{\prime}_1=k^{\prime}_1+n\frac{2\pi}{N}=k^{\prime}_1+n\frac{2\pi}{N}\nu \, mod \, BZ$ as well as $p_1=k_1+n\frac{2\pi}{N}=k_1+n\frac{2\pi}{N}\nu \, mod \, BZ$ with $n^{\prime},n\in\{0,...,N-1\}$. This matrix representation with discrete indices $n^{\prime}$ and $n$ is called Harper representation. With the quantization of the magnetic field in the form $B=2\pi\frac{\nu}{N}$ it can then be seen that $\pm B$ induces a shift in the Harper representation by $n\to n\pm 1$. We emphasize that the size of the matrices in Harper representation takes as input $\nu$ and $N$ which are uniquely determined by the external homogeneous magnetic field.

\section{Weyl symbol of operator}
\label{AppC}

We will derive the Ward-Takahashi identity of a fermionic quantum field theory with spin $\frac{1}{2}$ in the case of non-homogeneity in $D$ spacetime dimensions. This non-homogeneity may, e. g., be due to an external electromagnetic or gravitational field. The usual derivation of the identity in the homogeneous case is performed in the momentum space representation of the theory. In the non-homogeneous case the momentum space fermion propagator is no longer diagonal. We will derive the Ward-Takahashi identity both in momentum representation and using the Wigner-Weyl calculus. We start with the general definitions of basic quantities, derive the momentum representation and then the Wigner transformation of the Ward-Takahashi identity and reduce our result for the inhomogeneous setup to that in the homogeneous case. We will use the system of units with $\hbar =1$.

The Wigner-Weyl calculus establishes a one-to-one correspondence between a quantum mechanical theory defined on a Hilbert space and its reformulation in terms of functions on phase space. An operator $\hat{O}$ on Hilbert space, being a function of the position operator $\hat{x}$ and the momentum operator $\hat{p}$, is associated with a phase space function $O_W(p,x)$, being a function of space $x$ and momentum $p$, by the definition

\begin{align}
	\hat{O}=\int \frac{d^Dk}{(2\pi )^D}\frac{d^Dp}{(2\pi )^D}d^Dyd^DxO_W(p,x)e^{i(k(x-\hat{x})+y(p-\hat{p}))}.
\end{align}

A wave function on Hilbert space is represented in bra-ket notation by $|\Psi\rangle$ with configuration space representation $\Psi (x)=\langle x|\Psi\rangle$. We would like to relate the phase space function $O_W(p,x)$ to the configuration space representation of $\hat{O}$ denoted by $O(x_1,x_2)=\langle x_1|\hat{O}|x_2\rangle$ as well as to its momentum space representation denoted by $\tilde{O}(p_1,p_2)=\langle p_1|\hat{O}|p_2\rangle$. The configuration and momentum space representations are related by

\begin{align}
	O(x_1,x_2)=\int\frac{d^Dp_1}{(2\pi )^{\frac{D}{2}}}\frac{d^Dp_2}{(2\pi )^{\frac{D}{2}}}e^{ip_1x_1}\tilde{O}(p_1,p_2)e^{-ip_2x_2}
\end{align}

In order to achieve this we calculate $\langle z|\hat{O}|\Psi\rangle$. Applying the identities

\begin{align}
	\langle z|e^{ik\hat{x}}|\Psi\rangle =e^{ikz}\langle z|\Psi\rangle =e^{ikz}\Psi (z),\,\,\,\, \langle z|e^{-iy\hat{p}}|\Psi\rangle =e^{-y\partial_z}\langle z|\Psi\rangle =e^{-y\partial_z}\Psi (z) =\Psi (z-y)
\end{align}

as well as the Baker-Campbell-Hausdorff relation

\begin{align}
	e^{-i(k\hat{x}+y\hat{p})}=e^{-ik\hat{x}}e^{-iy\hat{p}}e^{i\frac{ky}{2}}
\end{align}

we obtain

\begin{align}
	\nonumber \langle z|\hat{O}|\Psi\rangle =&\langle z|\int \frac{d^Dk}{(2\pi )^D}\frac{d^Dp}{(2\pi )^D}d^Dyd^DxO_W(p,x)e^{i(k(x-\hat{x})+y(p-\hat{p}))}|\Psi\rangle\\
	\nonumber =&\int \frac{d^Dk}{(2\pi )^D}\frac{d^Dp}{(2\pi )^D}d^Dyd^DxO_W(p,x)e^{i(kx+yp+\frac{ky}{2})}e^{-ikz}\Psi (z-y)\\
	\nonumber =&\int \frac{d^Dp}{(2\pi )^D}d^Dyd^DxO_W(p,x)\delta^D(x+\frac{y}{2}-z)e^{iyp}\Psi (z-y)\\
	=&\frac{1}{(2\pi )^D}\int d^Dpd^DyO_W(p,z-\frac{y}{2})e^{ipy}\Psi (z-y).
\end{align}

A comparison with the configuration space relation $\langle z|\hat{O}|\Psi\rangle =\int d^DxO(z,x)\Psi (x)$ for $x=z-y$ implies

\begin{align}
	O(x,y)=\frac{1}{(2\pi )^D}\int d^DpO_W(p,\frac{x+y}{2})e^{ip(x-y)}
\end{align}

An inversion may be performed by a change of variables $R=\frac{x+y}{2}$ and $r=x-y$ followed by a Fourier transformation with respect to $r$. It reads

\begin{align}
	O_W(p,R)=\int d^DyO(R+\frac{y}{2},R-\frac{y}{2})e^{-ipy}
\end{align}

The analogous relation for the momentum space representation may be obtained similarly and reads

\begin{align}
	\tilde{O}(p,k)=\frac{1}{(2\pi )^D}\int d^DxO_W(\frac{p+k}{2},x)e^{i(k-p)x}
	\label{momentumww}
\end{align}

with inverse

\begin{align}
	O_W(p,R)=\int d^Dk\tilde{O}(p+\frac{k}{2},p-\frac{k}{2})e^{ikR}.
\end{align}

\section{Topological invariance of persistent current}

\label{AppA}

The proof applies to both the free and interacting theories alike, which is why we omit the subscripts used previously to distinguish them. Consider a variation of the averaged current 

\begin{align}
	\nonumber\delta J^k=&-\frac{1}{T L^2}\int d^3x\frac{d^3p}{(2\pi )^3}\delta Tr(G_W(x,p)\star\partial_{p_k} Q_W(x,p))\\
	\nonumber=&-\frac{1}{T L^2}\int d^3x\frac{d^3p}{(2\pi )^3}Tr(\delta G_W(x,p)\star\partial_{p_k} Q_W(x,p)+G_W(x,p)\star\partial_{p_k} \delta Q_W(x,p))\\
	\nonumber=&-\frac{1}{T L^2}\int d^3x\frac{d^3p}{(2\pi )^3}Tr(-G_W(x,p)\star \delta Q_W(x,p)\star G_W(x,p)\star\partial_{p_k} Q_W(x,p)+G_W(x,p)\star\partial_{p_k} \delta Q_W(x,p))\\
	\nonumber=&-\frac{1}{T L^2}\int d^3x\frac{d^3p}{(2\pi )^3}Tr(-\delta Q_W(x,p)\star G_W(x,p)\star\partial_{p_k} Q_W(x,p)\star G_W(x,p)+G_W(x,p)\star\partial_{p_k} \delta Q_W(x,p))\\
	\nonumber=&-\frac{1}{T L^2}\int d^3x\frac{d^3p}{(2\pi )^3}Tr(\delta Q_W(x,p)\star \partial_{p_k}G_W(x,p)+G_W(x,p)\star\partial_{p_k} \delta Q_W(x,p))\\
	\nonumber=&-\frac{1}{T L^2}\int d^3x\frac{d^3p}{(2\pi )^3}\partial_{p_k}Tr(\delta Q_W(x,p)\star G_W(x,p))\\
	=&0
\end{align}

where we used the cyclicity of the trace and the periodicity in momentum space.\\

\section{Ward identity in terms of Wigner transformed Green functions}
\label{AppB}

We will give here the generalized Ward-Takahashi identity (following method proposed in the textbook by  Weinberg). The vacuum state of the system is denoted by $|vac\rangle$. Our starting point is the Green function for the electric current $J^{\mu}(x)$ together with a Heisenberg picture Dirac field $\Psi_n(y)$ of electric charge $q$ and its covariant adjoint $\bar{\Psi}_m(z)$. We define the electromagnetic vertex function $\Gamma^{\mu}$ for an inhomogeneous system by 

\begin{align}
	\nonumber &\int d^Dxd^Dyd^Dze^{-ipx}e^{-ik_1y}e^{ik_2z}\langle vac|T\{ J^{\mu}(x)\Psi_n(y)\bar{\Psi}_m(z)\}|vac\rangle\\
	&\equiv -iq \int d^Dl_1d^Dl_2{\cal G}_{nn^{\prime}}(k_1,l_1)\Gamma^{\mu}_{n^{\prime}m^{\prime}}(l_1,l_2|p){\cal G}_{m^{\prime}m}(l_2,k_2)
	\label{vertexfunction}
\end{align}

with time ordering symbol $T$ and momentum space Dirac propagator

\begin{align}
	-i{\cal G}_{nm}(k,l)=\int d^Dyd^Dz\langle vac|T\{ \Psi_n(y)\bar{\Psi}_m(z)\}|vac\rangle .
\end{align}

With the canonical momentum conjugate $\Pi$ for $\Psi$ as well as the expression for the current $J^{\mu}$ (for one charged particle species)

\begin{align}
	\Pi (x)=\frac{\partial\mathcal{L}}{\partial (\partial_0\Psi (x))},\,\,\,\, J^{\mu}(x)=-i\frac{\partial\mathcal{L}}{\partial (\partial_{\mu}\Psi (x))}\Psi (x),
\end{align}

the canonical equal time commutation relation 

\begin{align}
	[\Psi_m (\bold{x},t),\Pi_n (\bold{y},t)]=i\delta^{D-1}(\bold{x}-\bold{y})\delta_{mn}
\end{align}

as well as the commutation relations

\begin{align}
	[J^{0}(\bold{x},t),\Psi (\bold{y},t)]=-q\Psi (\bold{y},t)\delta^{D-1}(\bold{x}-\bold{y}),\,\,\,\, [J^{0}(\bold{x},t),\bar{\Psi} (\bold{y},t)]=q\bar{\Psi} (\bold{y},t)\delta^{D-1}(\bold{x}-\bold{y})
\end{align}

we can derive a relation between the vertex function $\Gamma^{\mu}$ and the Dirac propagator $S$ as follows. Use the previous relations to obtain

\begin{align}
	\nonumber &\frac{\partial}{\partial x^{\mu}}T\{ J^{\mu}(x)\Psi_n(y)\bar{\Psi}_m(z)\}\\
	\nonumber &=T\{ \partial_{\mu}J^{\mu}(x)\Psi_n(y)\bar{\Psi}_m(z)\}+\delta (x^0-y^0)T\{ [J^0(x),\Psi_n(y)]\bar{\Psi}_m(z)\} +\delta (x^0-z^0)T\{ \Psi_n(y)[J^0(x),\bar{\Psi}_m(z)]\}\\
	&=-q\delta^D(x-y)T\{ \Psi_n(y)\bar{\Psi}_m(z)\}+q\delta^D(x-z)T\{ \Psi_n(y)\bar{\Psi}_m(z)\} .
\end{align}

Contraction of (\ref{vertexfunction}) with $p_{\mu}$ and using the obtained quantum current conservation law then leads to (where from now on we omit spin indices)

\begin{align}
	\int d^Dl_1d^Dl_2p_{\mu}{\cal G}(k_1,l_1)\Gamma^{\mu}(l_1,l_2|p){\cal G}(l_2,k_2)=i{\cal G}(p+k_1,k_2)-i{\cal G}(k_1,k_2-p).
	\label{keyrelation}
\end{align}

An alternative derivation may be given by making use of the Schwinger-Dyson equations and the relation between the conserved current and the variation of the action under the global symmetry corresponding to charge conservation.\\
To proceed act with $\int d^Dk_1d^Dk_2{\cal Q}(p_1,k_1)$ to the left and with ${\cal Q}(k_2,p_2)$ to the right of (\ref{keyrelation}). This gives

\begin{align}
	\nonumber &p_{\mu}\Gamma^{\mu}(p_1,p_2|p)\\
	\nonumber &=\int d^Dl_1d^Dl_2p_{\mu}\delta^D(p_1-l_1)\Gamma^{\mu}(l_1,l_2|p)\delta^D(l_2-p_2)\\
	\nonumber &=\int d^Dl_1d^Dl_2d^Dk_1d^Dk_2p_{\mu}{\cal Q}(p_1,k_1){\cal G}(k_1,l_1)\Gamma^{\mu}(l_1,l_2|p){\cal G}(l_2,k_2){\cal Q}(k_2,p_2)\\
	\nonumber &=i\int d^Dk_1d^Dk_2({\cal Q}(p_1,k_1){\cal G}(p+k_1,k_2){\cal Q}(k_2,p_2)-{\cal Q}(p_1,k_1){\cal G}(k_1,k_2-p){\cal Q}(k_2,p_2))\\
	\nonumber &=i(\int d^Dk_1{\cal Q}(p_1,k_1)\delta^D(p+k_1-p_2)-\int d^Dk_2\delta^D(p_1-k_2+p){\cal Q}(k_2,p_2))\\
	&=i({\cal Q}(p_1,p_2-p)-{\cal Q}(p_1+p,p_2)).
\end{align}

Therefore we have

\begin{align}
	p_{\mu}\Gamma^{\mu}(p_1,p_2|p)=i({\cal Q}(p_1,p_2-p)-{\cal Q}(p_1+p,p_2)).
	\label{wardtakamomentum}
\end{align}

This is the generalized Ward-Takahashi identity in momentum space representation.\\
We may now insert the relation (\ref{momentumww}) for all momentum space quantities into (\ref{wardtakamomentum}). This yields

\begin{align}
	\frac{1}{(2\pi )^D}\int d^Dxp_{\mu}\Gamma^{\mu}_W(\frac{p_1+p_2}{2},x|p)e^{i(p_2-p_1)x}
	=&\frac{i}{(2\pi )^D}\int d^Dx({\cal Q}_W(\frac{p_1+p_2-p}{2},x)-{\cal Q}_W(\frac{p_1+p_2+p}{2},x))e^{i(p_2-p_1-p)x}.
\end{align}

A Fourier transformation with respect to $p_2-p_1$ together with a renaming of the momentum variables leads to

\begin{align}
	p_{\mu}\Gamma^{\mu}_W(P,x|p)=i({\cal Q}_W(P-\frac{p}{2},x)-{\cal Q}_W(P+\frac{p}{2},x))e^{-ipx}
	\label{wardtakaww}
\end{align}

which is the generalized Ward-Takahashi identity in phase space representation.\\
Consider now the limit $p^{\mu}\to 0$. A first order Taylor expansion on the right hand side of (\ref{wardtakaww}) followed by the momentum limit implies the generalized Ward identity

\begin{align}
	\Gamma^{\mu}_W(P,x|0)=-i\frac{\partial}{\partial p_{\mu}}{\cal Q}_W(p,x)|_{p=P}.
\end{align}

We would finally like to reduce the results of (\ref{wardtakamomentum}) and (\ref{wardtakaww}) to the homogeneous case. In order to achieve this it is best to start from the momentum space representation of the identity given in (\ref{wardtakamomentum}). In the homogeneous case the momentum space Dirac operator becomes diagonal ${\cal G}(k,l)\to {\cal G}(k)\delta^D(k-l)$ and the vertex function reduces to $\Gamma^{\mu}(p_1,p_2|p)\to \Gamma^{\mu}(p_1,p_2)\delta^D(p+p_1-p_2)$. This leads to

\begin{align}
	\nonumber p_{\mu}\Gamma^{\mu}(p_1,p_2)\delta^D(p+p_1-p_2)=&i({\cal Q}(p_1)\delta^D(p_1-p_2+p)-{\cal Q}(p_1+p)\delta^D(p_1+p-p_2))\\
	=&i({\cal Q}(p_1)-{\cal Q}(p_2)).
\end{align}

Performing an integral over the momentum $p$ implies the homogeneous Ward-Takahashi identity in momentum space

\begin{align}
	\nonumber (p_2-p_1)_{\mu}\Gamma^{\mu}(p_1,p_2)=&i({\cal Q}(p_1)-{\cal Q}(p_2)).
\end{align}

The homogeneous Ward identity follows by taking the momentum coincidence limit $p_2\to p_1$ or vice versa. After a first order Taylor expansion on the right hand side this limit leads to

\begin{align}
	\Gamma^{\mu}(p,p)=-i\frac{\partial}{\partial p_{\mu}}{\cal Q}(p).
\end{align}

\end{document}